\documentclass[journal]{IEEEtran}
\usepackage{graphicx,bm}
\usepackage{amsmath}
\usepackage{color}
\usepackage[dvips]{epsfig}
\usepackage{graphicx}
\usepackage{calligra}
\allowdisplaybreaks[4]
\usepackage{amsfonts}
\usepackage{amssymb}
\usepackage{amsthm}
\usepackage[numbers,sort&compress]{natbib}
\usepackage{mathrsfs}
\usepackage{txfonts}
\usepackage{dsfont}
\usepackage[T1]{fontenc}
\usepackage[latin9]{inputenc}
\usepackage{units}
\usepackage{esint}
\usepackage{graphicx,bm}
\usepackage{amsmath}
\usepackage{amsfonts}
\usepackage{color}
\usepackage{amssymb}
\usepackage{amsthm}
\usepackage{mathrsfs}
\usepackage{txfonts}
\usepackage{dsfont}
\usepackage[T1]{fontenc}
\usepackage[latin9]{inputenc}
\usepackage{units}
\usepackage{esint}
\usepackage{float}
\usepackage{graphicx,bm}
\usepackage{stfloats}
\usepackage{amsmath}
\usepackage{amsfonts}
\usepackage{color}
\usepackage{amssymb}
\usepackage{amsthm}
\usepackage{mathrsfs}
\usepackage{bm}
\usepackage{fancyhdr}
\usepackage{epstopdf}
\usepackage{txfonts}
\usepackage{dsfont}
\usepackage{multirow}
\usepackage{amsmath, amsthm, amssymb}
\usepackage{comment}
\usepackage{latexsym}
\usepackage{bbm}
\newcounter{bla}
\begin{document}
%\captionsetup[figure]{name={Fig.},labelsep=period}
\title{On the Impact of Unknown Signals in Passive Radar with Direct Path and Reflected Path Observations}

\IEEEoverridecommandlockouts
\author{Yicheng Chen, \IEEEmembership{Student Member,~IEEE}, and Rick S. Blum,
\IEEEmembership{Fellow,~IEEE}
\thanks{The work of Yicheng Chen and R. S.
Blum was supported by the National Science Foundation under Grant
No. ECCS-1405579. This material is based upon work partially supported by the U. S. Army Research Laboratory and the U. S. Army Research Office under grant number W911NF-17-1-0331. }
\thanks{
Yicheng Chen and R. S. Blum are with Lehigh University, Bethlehem, PA 18015 USA
(email: yic917@lehigh.edu, rblum@eecs.lehigh.edu). }} \IEEEoverridecommandlockouts

\maketitle

\begin{abstract}
We derive the closed form Cramer-Rao bound (CRB) expressions for joint estimation of time delay and Doppler shift with unknown signals with possibly known structure. The results are especially useful for passive radar where direct path and reflected path signals are present. Time delay and Doppler shift estimation is an important fundamental tool in signal processing which has received extensive study for cases with known transmitted signals, but little study for unknown transmitted signals. The presented results generalize previous results for known transmitted signals and show how many looks from the direct path and the reflected path we need to derive an accurate joint estimation of time delay and Doppler shift. After analysis under a simple common  signal-to-clutter-plus-noise ratio (SCNR) model with separated direct and reflected path signals, white clutter-plus-noise and line of sight propagation, extensions to cases with different direct and reflected path SCNRs, correlated clutter-plus-noise, nonseparated direct and reflected path signals and multipath propagation are discussed to support the utility of the CRB with unknown signals.
\end{abstract}


\IEEEpeerreviewmaketitle

\section{Introduction}

%{\bf I can fix this later, but let's start with a few selected references that I will suggest separately in an email}
%  In recent years, passive radars employing existing wireless communications have received extensive research attention \cite{Bournaka:2017}--\cite{Zheng:2017}. Existing signals from frequency modulation (FM) radio \cite{Fu:2017}, digital video broadcasting terrestrial (DVB-T) \cite{Martelli:2017}, digital audio broadcasting (DAB) \cite{Schupbach:2017}, WiFi \cite{Javed:2017}, global system for mobile communications (GSM) \cite{Tabassum:2017}, worldwide interoperability for microwave access (WiMAX)  \cite{Higgins:2016} and  Long Term Evolution (LTE) \cite{Abdullah:2016} systems have been studied to provide illuminators of opportunity in passive radar systems.
%\cite{bell1995performance} \cite{bell1996explicit} \cite{xu2001performance} \cite{xu2006performance} \cite{kozick2004source} \cite{sadler2007ziv} \cite{chiriac2009performance} \cite{sadler2010ziv} \cite{liu2010ziv} \cite{chiriac2010ziv} \cite{li2007mimo} \cite{tabrikian2006barankin}

The topic of time delay and Doppler shift estimation continues to attract attention \cite{Zhang:2016,Bell:1997, bell1996explicit,kozick2004source,sadler2007ziv,chiriac2009performance,liu2010ziv,chiriac2010ziv,li2007mimo,tabrikian2006barankin,xu2006performance,van2007bayesian,Zheng:2017,Shi:2016,Exponential:2017}, since it is recognized to be a basic problem of significant interest in radar, communications and related sensor signal processing systems.  Early work has built the foundation for four decades of research on time delay and Doppler shift estimation, see for example \cite{Lank:1973,Knapp:1976,Axline:1977,Wax:1982,Friedlander:1984}.
%  which continues to be of great interests today \cite{riedlander:2012,Shames:2013,Sajjadieh:2015,Zhang:2016,Exponential:2017}.  Recent publications continue to document the interest in these topics as the following references demonstrate.  A high-resolution algorithm based on compressive sensing for passive radars that makes use of orthogonal frequency-division multiplexing (OFDM) communication signals for joint time delay and Doppler shift estimation is proposed in \cite{Zheng:2017}. A study of joint estimation of the direction of arrival (DOA), Doppler information, and direction of departure (DOD) based on compressive sensing and time reversal is presented in \cite{Sajjadieh:2015}. The joint time delay and Doppler shift estimation problem in passive radar systems in the presence of direct-path interference is studied  in \cite{Zhang:2016}. Joint estimation of delay and Doppler in the presence of multiple targets in radar applications based on an exponential ambiguity function and wavelet denoising method is given in \cite{Exponential:2017}.
The recent increased attention on passive radar systems is noticeable \cite{hack2014detection,hack2014centralized,liu2014two,cui2014target,wang2016canonical,zhang2017multistatic,palmer2013dvb,Zheng:2017,chabriel2017adaptive,zaimbashi2017forward}. We attribute this to several advantages, including smaller size, less detectable radar operation, more portability and lower cost over traditional radar systems, referred to as active radar systems.

An informative way to evaluate the estimation performance in a radar system is to employ an achievable lower bound on the estimation error. The Cramer-Rao bound (CRB) is a widely used lower bound on the variance (or mean square error) of  all unbiased estimators which is achievable using maximum likelihood estimators under mild conditions.  The CRB is regarded as an important benchmark of performance in radar systems \cite{Shi:2016,Friedlander:1984,Javed:2017,gogineni2014cramer,he2014significant}. The CRB for estimating the time delay and Doppler shift of the target in passive radar systems has been calculated in \cite{Shi:2016,Javed:2017,gogineni2014cramer,he2014significant} under the assumption that the transmitted signal can be perfectly estimated, so the transmitted signal is assumed known. However, in practice, the exact transmitted signals from non-cooperative illuminators of opportunity in passive radar system are typically unknown by the passive radar system which is a topic that has not seen much investigation \textcolor[rgb]{0.00,0.00,0.00}{for estimation of target parameters}.

The impact of unknown signals on passive radar performance is of critical importance.  In practice we typically have both direct path and reflected path observations over certain time periods, and the direct path delay and Doppler may be known or previously estimated such that it can be removed.  Thus a canonical problem is to estimate the parameters from both a delayed and Doppler shifted version of the signal along with a zero delayed and zero Doppler shifted version of the signal.  The zero delayed and zero Doppler shifted version of the signal comes from the direct path, while the delayed and Doppler shifted version of the signal comes from the reflected path. In different systems, the direct and reflected path may be separated by antenna array processing. Further in some cases, we get multiple looks at both the direct path and the reflected path signals and we would like to know how these multiple looks impact our performance, along with all the other parameters.  The multiple looks could come from observations from closely spaced antenna array elements which see different noise observations but similar signals, delays and Doppler frequencies.
%  {\bf Is this interesting? Should we consider it here or say it is out of scope but could be investigated}.----Blum
  While other bounds can be employed, the CRB seems to be the simplest and most studied lower bound.  Thus, using the CRB seems a proper first step in this relatively unstudied direction that attempts to evaluate parameter estimation performance with unknown signals with either unknown or known structure.  Knowing the relationship between the CRBs with unknown and  known signals allows designers to understand the loss and decide if they should increase the number of looks or modify something else to close this gap. We also consider the impact of knowing the signal structure which is also very important and can often be exploited in passive radar.

We found only one recent paper \cite{Decarli:2014} which considered the performance of time delay estimation with unknown deterministic signals. While \cite{Decarli:2014} is an interesting and useful paper, the Ziv-Zakai bound is employed in \cite{Decarli:2014} after an unjustified replacement of the likelihood ratio test required by the Ziv-Zakai bound with a generalized likelihood ratio test. This modification destroys the validity of the Ziv-Zakai bound such that it is no longer known to be a bound or to have any known relationship to the actual estimation performance so that the provided results are not guaranteed to be meaningful. Further, \cite{Decarli:2014} does not provide simple closed-form expressions, other than those involving a very complicated integral which almost always requires numerical evaluation, which limits insight. We, however, provide justified bounds\textcolor[rgb]{0.00,0.00,0.00}{\footnote{Achievable by maximum likelihood estimator with a sufficient number of obsevations.}} and simple closed-form expressions which are not available in \cite{Decarli:2014}.
%However, the results in \cite{Decarli:2014} have not been justified. Besides, there is little work on the impact of unknown signals for time delay and Doppler shift estimation. And it is unknown how much the estimation performance will change when we have different number of observations of unknown signals. Such information would be significantly useful in passive radar system for designers to evaluate whether they need more observations to achieve required estimation performance.

In this paper, we consider the impact of unknown signals in passive radar with direct path and reflected path observations, but we employ canonical models, with the hope that these results might be adopted in other applications with
unknown signals. We derive a closed-form expression for the CRB for joint time delay and Doppler shift estimation for cases with unknown signals with either unknown or known signal structure and possibly multiple looks at the direct path and reflected path returns. We explicitly consider known structure signals consisting of amplitude modulated pulse trains. After the main analysis is described for a simple model, we discuss extensions to more complicated models.
  The main contributions of this paper are:\\
  1. For the case of unknown signal structure and a simplified model, closed-form expressions of the Fisher information matrix (FIM) and the CRB for joint time delay and Doppler shift estimation with unknown signals are derived based on possible multiple looks at the direct path and reflected path returns. The relationship between the unknown signals CRB and the known signals CRB is obtained.  It is shown that the unknown signals CRBs for delay and Doppler shift are each the product of the corresponding known signal CRB multiplied by a simple factor that depends  on the number of looks. With a single look  from the reflected path, which includes the delay and Doppler shift,  and a sufficiently large number of looks at the direct path, which does not include the delay and Doppler shift,  the CRB for unknown signals approaches that for known signals.  Thus, the observations can be used to accurately estimate the unknown signal. Further, for a sufficiently large number of looks at both the direct and reflected paths, the CRBs can be driven to zero. \\
  2. Similar expressions are provided for the case where the signal has some known structure such that it can be described by some known expressions with some unknown parameters representing information embedded into the signals.  A specific case using amplitude modulation is used to make these ideas concrete.  The estimation  performance is shown to improve when the known structure is acknowledged.  \\
  3. Extensions to more complicated models with different direct and reflected path SCNRs, correlated clutter-plus-noise, nonseparated direct and reflected path signals, and multipath returns are described and detailed solutions are provided or outlined.

The paper is organized as follows. Joint time delay and Doppler shift estimation with known signals is discussed in Section \uppercase\expandafter{\romannumeral 2}. The closed-form expressions of the CRB for joint time delay and Doppler shift estimation with unknown signals are developed in Section \uppercase\expandafter{\romannumeral 3}. In Section \uppercase\expandafter{\romannumeral 4}, we derive the closed-form expressions of the CRB for joint estimation with known signal structure. Numerical examples provide the CRB for joint estimation with unknown signals with either unknown or known structure in Section \uppercase\expandafter{\romannumeral 5}. In Section \uppercase\expandafter{\romannumeral 6}, extensions to the observation model are considered. Finally, Section \uppercase\expandafter{\romannumeral 7} concludes the paper.

Throughout this paper, the notation for transpose is $T$, while the symbol $\left|\, \right|$ denotes the norm. Bold lower case letters are used to denote column vectors, and bold upper case letters denote matrices. Let ${\bm{A}}_{i,j}$ denote the element in the $i$-th row and $j$-th column of the matrix $\bm A$, $\bm{1}$ denote an identity matrix and $\bm{0}$ denote the all zero matrix. $\mathbb{E}$ denotes the expectation operator. $\text{Tr}\left(
\cdot \right)$ denotes the trace of a matrix, $\otimes$ represents
the Kronecker product, and $\text{vec}(\cdot)$
denotes the vectorizing operator which stacks the columns of
a matrix in a column vector.

\section{Joint estimation with a known signal}

Consider the reflected path signal with unknown time delay $\tau_0$ and Doppler shift $f_0$ for a completely known finite support narrow-band complex baseband \textcolor[rgb]{0.00,0.00,0.00}{transmitted} signal $s(t)$.  Suppose we take discrete-time samples with a reasonably small $\Delta$ to obtain the baseband observations %\cite{kay}
\begin{eqnarray}
 x(n\Delta) = s(n\Delta-\tau_0){e^{j2\pi {f_0} n\Delta }} + w(n\Delta) \label{obs}
\end{eqnarray}
for $n = 0,1,\ldots, N-1$ with $ \tau_0 = n_0 \Delta$. If $ s(n\Delta) $ is zero for $n < 0 $ and $ n > (M-1)\Delta $, then  $ x(n\Delta) = $
\begin{eqnarray}
 \left\{
                \begin{array}{ll}
                  w(n\Delta)  & \mbox{ if } 0 \leq n \leq n_0 - 1 \\
                  s(n\Delta-\tau_0){e^{j2\pi {f_0} n\Delta }} + w(n\Delta) & \mbox{ if }  n_0 \leq n \leq n_0 + M - 1
                  \\
                  w(n\Delta) & \mbox{ if }  n_0 + M \leq n \leq N - 1
                \end{array}
              \right. .
\nonumber
%\label{sigdur}
\end{eqnarray}
The observations in (\ref{obs}) are often called the reflected-path observations in passive radar \textcolor[rgb]{0.00,0.00,0.00}{and they} can be obtained by pointing a directional antenna \textcolor[rgb]{0.00,0.00,0.00}{in} the target direction. Assuming independent and identically distributed (iid) complex Gaussian zero mean and variance $\sigma_w^2$ clutter-plus-noise
samples $ w(n\Delta), n=0,\ldots,N-1 $, then we use the JCRB to denote the CRB \textcolor[rgb]{0.00,0.00,0.00}{for}
any unbiased (zero mean) joint time delay and Doppler shift estimation ($ \hat{\tau_0}$, $ \hat{f_0}$) based on the observations
$( x(0), x(\Delta), \ldots, x((N-1)\Delta)^T$
which implies \textcolor[rgb]{0.00,0.00,0.00}{(see Appendix A)}
\begin{small}
\begin{eqnarray} \label{tauknown}
var(\hat{\tau_0}) &\ge& JCR{B_{{\tau _0}}}
 \nonumber \\ &=&
\frac{{\sigma _w^2\sum\limits_{n = 0}^{M - 1} {{{(t + {\tau _0})}^2}{{\left| {s(t)} \right|}^2}}\biggr{|}_{t=n \Delta} }}{{2\left( {\sum\limits_{n = 0}^{M - 1} {{{\left| {\frac{{\partial s(t)}}{{\partial t}}} \right|}^2}\biggr{|}_{t=n \Delta}} \sum\limits_{n = 0}^{M - 1} {{{(t + {\tau _0})}^2}{{\left| {s(t)} \right|}^2}\biggr{|}_{t=n \Delta}}  - {\eta ^2}} \right)}}
\end{eqnarray}
\end{small}
and
\begin{small}
\begin{eqnarray}\label{fknown}
 var(\hat{f_0})&\ge& JCR{B_{{f_0}}}
 \nonumber \\ &=&
\frac{{\sigma _w^2\sum\limits_{n = 0}^{M - 1} {{{\left| {\frac{{\partial s(t)}}{{\partial t}}} \right|}^2}}\biggr{|}_{t=n \Delta} }}{{8{\pi ^2}\left( {\sum\limits_{n = 0}^{M - 1} {{{\left| {\frac{{\partial s(t)}}{{\partial t}}} \right|}^2}\biggr{|}_{t=n \Delta}} \sum\limits_{n = 0}^{M - 1} {{{(t + {\tau _0})}^2}{{\left| {s(t)} \right|}^2}\biggr{|}_{t=n \Delta}}  - {\eta ^2}} \right)}}
\end{eqnarray}
\end{small}
where $\eta$ is defined as
\begin{eqnarray}\label{eta}
\eta  = \sum\limits_{n = 0}^{M - 1} {(t + {\tau _0})} ({s_I}(t)\frac{{\partial {s_R}(t)}}{{\partial t}} - {s_R}(t)\frac{{\partial {s_I}(t)}}{{\partial t}})\biggr{|}_{t=n \Delta}
\end{eqnarray}
and ${s_R}(t)$, $ {s_I}(t)$ are the real and imaginary parts of ${s}(t)$ respectively. We have treated $ \tau_0 $
as a continuous variable for convenience, but this is a reasonable approximation for sufficiently fast sampling \cite{kay}.
%If the signal $s(t)$ is an electromagnetic wave traveling at the speed $c$ %then the variance of an unbiased estimate of the range $ \hat{r_0}$
%between the transmitter and receiver obeys $ var(\hat{r_o}) \geq c^2 %CRB_{\tau_0} $.  Range estimation based on time-delay is fundamental for %locating the

%{\bf Let's get rid of the names (SLAJS) and (MLAJS) }

\section{Joint estimation with unknown signals}\label{methodwe}

Now assume that $s(t)$ is an unknown function of $t$ to model the case where the narrow-band \textcolor[rgb]{0.00,0.00,0.00}{transmitted} signal is unknown.  This is a problem of interest for passive radar. Under the same high sampling rate assumptions so that the approximation of continuous time delay $ \tau_0 $ is sufficiently accurate, then we can characterize the losses from not knowing the signal by calculating the CRB \textcolor[rgb]{0.00,0.00,0.00}{for} joint \textcolor[rgb]{0.00,0.00,0.00}{estimation} of the components of the parameters of the vector ${\bm\theta} = (  \tau_0, f_0,
{s_R}(0), {s_I}(0), {s_R}(\Delta), \ldots, {s_I}((M-1)\Delta)^T$. As common in passive radar, some direct path observations can be obtained by pointing directional antennas at the transmitter \cite{hack2014detection} and removing a known time delay. Let us assume that we augment the observations from the reflected path in (\ref{obs}) with
\begin{eqnarray}
 x_{d\ell}(n\Delta) &=& s(n\Delta) + \textcolor[rgb]{0.00,0.00,0.00}{w_{d\ell}(n\Delta)}   \label{refobs}
\end{eqnarray}
\textcolor[rgb]{0.00,0.00,0.00}{for $n = 0,1,\ldots, N-1, \ell = 1, \ldots, L$,} which we call reference observations, to help us estimate the signal samples.  We call the case without reference observations  the $L=0$ case. Assuming $L>0$, then the observations in (\ref{refobs}) provide $L$ extra looks at the undelayed and nonshifted signals.
Note that combining (\ref{obs}) with (\ref{refobs}) together, we \textcolor[rgb]{0.00,0.00,0.00}{obtain} $L$ looks at the zero delayed and zero Doppler shifted version of the signal
and one look at the delayed and Doppler shifted version of the signal.
All complex clutter-plus-noise samples in (\ref{obs}) and (\ref{refobs}) form an iid sequence \textcolor[rgb]{0.00,0.00,0.00}{(same model as $w$ in (\ref{obs}))}, and the real and imaginary parts of the signal samples ${s_R}(0), {s_I}(0), {s_R}(\Delta), \ldots, {s_I}((M-1)\Delta) $ are assumed to be deterministic unknowns.
%We call this the single look at joint signal (SLAJS) model since we only get one look at the time delayed and Doppler shifted signal.

\subsection{Generalization \textcolor[rgb]{0.00,0.00,0.00}{of} the Model}\label{method}

%{You named the observed reference signals the same in both cases? call direct path $x_{d\ell}$ and reflected  $x_{r\ell}$ ?
%so you can define ${\bf x}$ - see my changes }

Suppose we generalize the model such that we get $P \geq 1 $ looks at the delayed and Doppler shifted signal as opposed to the $P=1$ case in (\ref{obs}).  Then we replace (\ref{obs}) with
\begin{eqnarray}
 x_{r\ell}(n\Delta) &=& s(n\Delta-\tau_0){e^{j2\pi {f_0} n\Delta }} + \textcolor[rgb]{0.00,0.00,0.00}{w_{r\ell}(n\Delta)}\label{changeobs}
\end{eqnarray}
\textcolor[rgb]{0.00,0.00,0.00}{for $n = 0,\ldots, N-1$, $\ell = 1, \ldots, P$,} and we augment these observations with the $L$ looks from (\ref{refobs}).
Again, all complex clutter-plus-noise samples in (\ref{changeobs}) form an iid sequence \textcolor[rgb]{0.00,0.00,0.00}{(same model as $w$ in (\ref{obs}))} and the real and imaginary parts of the signal samples ${s_R}(0), {s_I}(0), {s_R}(\Delta), \ldots, {s_I}((M-1)\Delta) $ are assumed to be deterministic unknowns.
Note that the $(i,j)$th entry of the FIM in this multiple parameter case can be computed as \cite{kay}
\begin{eqnarray}\label{FIMkay}
{\bm I}{(\bm\theta )_{i,j}} =  -\mathbb E\left[ {\frac{{{\partial ^2}\ln p({\bm{x}};{\bm\theta} )}}{{\partial {\bm\theta _i}\partial {\bm\theta _j}}}} \right]
\end{eqnarray}
where \textcolor[rgb]{0.00,0.00,0.00}{the log of} the probability density function (pdf) $p(\bm{x};\bm\theta )$ of $ \bm{x} = (  x_{r1}(0), \ldots, x_{r1}((N-1)\Delta),x_{r2}(0), \ldots, x_{rP}((N-1)\Delta), x_{d1}(0), \ldots, x_{d1}((N-1)\Delta), x_{d2}(0), \ldots, x_{dL}((N-1)\Delta))^T $ \textcolor[rgb]{0.00,0.00,0.00}{is}
\begin{eqnarray}
\ln p(\bm{x};\bm{\theta}) \propto \frac{-1}{\sigma_w^2} \Bigg( \sum_{l=1}^P \sum_{n=0}^{N-1} \Big| x_{rl}(n\Delta) - s(n\Delta - \tau_0)e^{j2\pi f_0 n\Delta}\Big|^2 \nonumber  \\
+\sum_{l=1}^{L} \sum_{n=0}^{N-1} \Big| x_{dl}(n\Delta) - s(n\Delta)\Big|^2 \Bigg).
\end{eqnarray}

The FIM for estimating $\bm\theta$ in this case is defined as
\begin{align}\label{FIMbrief}
{\bm{I}}(\bm\theta ) = \left[ {\begin{array}{*{20}{c}}
{{\bm{A}}}&{{\bm{B}}}\\
{\bm{B}^T}&{{\bm{C}}}
\end{array}} \right]
\end{align}
where the specific entries in the ${2 \times 2}$ symmetric matrix ${\bm{A}}$ in (\ref{FIMbrief}) are
\begin{small}
\begin{eqnarray}\label{A11}
 \bm{A}_{1,1} = -\mathbb E[\frac{{{\partial ^2}\ln p(\bm{x};\bm\theta)}}{{\partial \tau _0^2}}] =\frac{{2P}}{{\sigma _w^2}}\sum\limits_{n = 0}^{M - 1} {{{\left| {\frac{{\partial s(t)}}{{\partial {t}}}} \right|}^2}}\biggr{|}_{t=n \Delta},\\
 \bm{A}_{2,2}=  -\mathbb E[\frac{{{\partial ^2}\ln p(\bm{x};\bm\theta)}}{{\partial f_0^2}}] =P\frac{{8{\pi ^2}}}{{\sigma _w^2}}\sum\limits_{n = 0}^{M - 1} {{{(t+\tau_0)}^2}{{\left| {s(t)} \right|}^2}}\biggr{|}_{t=n \Delta},\label{A22}
\end{eqnarray}
\end{small}
and
\begin{small}
\begin{align}\label{A12}
\bm{A}_{1,2}=\bm{A}_{2,1}=-\mathbb E[\frac{{{\partial ^2}\ln p(\bm{x};\bm\theta)}}{{\partial {\tau _0}\partial {f_0}}}]=\frac{{4\pi P}}{{\sigma _w^2}}\eta %=\frac{{4\pi P}}{{\sigma _w^2}}\sum\limits_{n = 0}^{M - 1} {(t+\tau) \left[ {{s_I}(t)\frac{{\partial {s_R}(t)}}{{\partial {t}}} - {s_R}(t)\frac{{\partial {s_I}(t)}}{{\partial {t}}}} \right]}\biggr{|}_{t=n \Delta}
\end{align}
\end{small}where $\eta$ is defined in (\ref{eta}).
The specific entries in the ${2 \times 2M}$ matrix ${\bm{B}}$ in (\ref{FIMbrief}) are
\begin{equation}\label{B1j}
\bm{B}_{1,j}=
\begin{cases}
-\mathbb E[\frac{{{\partial ^2}\ln p(\bm{x};\bm\theta )}}{{\partial {s_R(n \Delta)}\partial {\tau_0}}}]=-\frac{{2P}}{{\sigma _w^2}}\frac{{\partial {s_R}(t)}}{{\partial {t}}}\biggr{|}_{t=n \Delta}&\mbox{if $ j=2n+1$}\\
-\mathbb E[\frac{{{\partial ^2}\ln p(\bm{x};\bm\theta )}}{{\partial {s_I(n \Delta)}\partial {\tau_0}}}] =-\frac{{2P}}{{\sigma _w^2}}\frac{{\partial {s_I}(t)}}{{\partial {t}}}\biggr{|}_{t=n \Delta}&\mbox{if $j=2n+2$},
\end{cases}
\end{equation}
and
\begin{equation}\label{B2j}
\bm{B}_{2,j}  =
\begin{cases}
-\mathbb E[\frac{{{\partial ^2}\ln p(\bm{x};\bm\theta )}}{{\partial {s_R(n \Delta)}\partial {f_0}}}] ={ - \frac{{4\pi P(t+\tau_0) {s_I}(t)}}{{\sigma _w^2}}}\biggr{|}_{t=n \Delta} &\mbox{if $ j=2n+1$}\\
-\mathbb E[\frac{{{\partial ^2}\ln p(\bm{x};\bm\theta )}}{{\partial {s_I(n \Delta)}\partial {f_0}}}] = {\frac{{4\pi P(t+\tau_0) {s_R}(t)}}{{\sigma _w^2}}}\biggr{|}_{t=n \Delta}&\mbox{if $j=2n+2$},
\end{cases}
\end{equation}
\textcolor[rgb]{0.00,0.00,0.00}{for $n = 0,\ldots, M-1$.}
The specific entries in the ${2M \times 2M}$ diagonal matrix ${\bm{C}}$ in (\ref{FIMbrief}) are
\begin{align}\label{Cjj}
\bm{C}_{j,j} = \frac{{2L + 2P}}{{\sigma _w^2}} \qquad &\mbox{if $j=1,2,...,2M$}.
\end{align}
The Schur complement relation \textcolor[rgb]{0.00,0.00,0.00}{\cite{petersen2008matrix}} has been used to derive
\begin{align}\label{FIMinverse}
\textcolor[rgb]{0.00,0.00,0.00}{{[I(\bm\theta )^{ - 1}]_{(\{1,2\},\{1,2\})}}}= {{{({{\bm{A}}} - {{\bm{B} }}{{{\bm{C}}}^{ - 1}}{\bm{B}^T})}^{ - 1}}}
\end{align}
where ${[I(\bm\theta )^{ - 1}]_{(\{1,2\},\{1,2\})}}$ denotes the sub-matrix of $I(\bm\theta )^{ - 1}$ which consists of the elements located in the first two rows and the first two columns. Using the expressions of elements in ${\bm{A}}$, ${\bm{B}}$ and ${\bm{C}}$ derived in (\ref{A11})--(\ref{Cjj}), we obtain
\begin{small}
\begin{align}\label{tfcontain}
&\bm{A} - \bm{B}{\bm{C}^{ - 1}}{\bm{B}^T} =\nonumber\\
& \left[ {\begin{array}{*{20}{c}}
{\frac{{LP}}{{L + P}}\frac{2}{{\sigma _w^2}}{\sum\limits_{n = 0}^{M - 1} {{{\left| {\frac{{\partial s(t)}}{{\partial t}}} \right|}^2}} }}\biggr{|}_{t=n \Delta}&{\frac{{4\pi \eta }}{{\sigma _w^2}}\frac{{LP}}{{L + P}}}\\
{\frac{{4\pi \eta }}{{\sigma _w^2}}\frac{{LP}}{{L + P}}}&{\frac{{LP}}{{L + P}}{\frac{{8{\pi ^2}}}{{\sigma _w^2}}\sum\limits_{n = 0}^{M - 1} {{{(t + {\tau _0})}^2}{{\left| {s(t)} \right|}^2}} }}\biggr{|}_{t=n \Delta}
\end{array}} \right]
\end{align}
\end{small}where $\eta$ is defined in (\ref{eta}).
Any unbiased joint estimate of time delay, Doppler shift and signal samples satisfies \cite{kay}
\begin{align}\label{taudef}
{\mathop{\rm var}} ({\hat{\tau}_0}) \ge JCR{B_{{\tau_0},{s}}} % MathType!End!2!1!
\end{align}
where \textcolor[rgb]{0.00,0.00,0.00}{\cite{petersen2008matrix}}
\begin{align}\label{tauands}
&JCR{B_{{\tau _0}, {s}}} = \left[I(\bm\theta )^{ - 1}\right]_{1,1}\nonumber\\
&= \left[(\bm{A} - \bm{B}{\bm{C}^{ - 1}}{\bm{B}^T})^{ - 1} \right]_{1,1}= \left[ \frac{adj(\bm{A} - \bm{B}{\bm{C}^{ - 1}}{\bm{B}^T})}{det(\bm{A} - \bm{B}{\bm{C}^{ - 1}}{\bm{B}^T})} \right]_{1,1}\nonumber\\
&= {\frac{{LP}}{{L + P}}{\frac{{8{\pi ^2}}}{{\sigma _w^2}}\sum\limits_{n = 0}^{M - 1} {{{(t + {\tau _0})}^2}{{\left| {s(t)} \right|}^2}}\biggr{|}_{t=n \Delta} }}\cdot \nonumber\\
&\frac{1}{{{\frac{{LP}}{{L + P}}{\frac{{8{\pi ^2}}}{{\sigma _w^2}}\sum\limits_{n = 0}^{M - 1} {{{(t + {\tau _0})}^2}{{\left| {s(t)} \right|}^2}\biggr{|}_{t=n \Delta}} }}{{\frac{{LP}}{{L + P}}\frac{2}{{\sigma _w^2}}{\sum\limits_{n = 0}^{M - 1} {{{\left| {\frac{{\partial s(t)}}{{\partial t}}} \right|}^2}\biggr{|}_{t=n \Delta}} }}}-{\left( {\frac{{4\pi \eta }}{{\sigma _w^2}}\frac{{LP}}{{L + P}}} \right)}^2}} \nonumber\\
&= \frac{{L + P}}{{LP}} \cdot \frac{{\frac{{\sigma _w^2}}{2}\sum\limits_{n = 0}^{M - 1} {{{(t + {\tau _0})}^2}{{\left| {s(t)} \right|}^2}}\biggr{|}_{t=n \Delta} }}{{\sum\limits_{n = 0}^{M - 1} {{{\left| {\frac{{\partial s(t)}}{{\partial t}}} \right|}^2}\biggr{|}_{t=n \Delta}\sum\limits_{n = 0}^{M - 1} {{{(t + {\tau _0})}^2}{{\left| {s(t)} \right|}^2}\biggr{|}_{t=n \Delta}} }  - {\eta ^2}}}\nonumber\\
&=\frac{L + P}{LP}JCR{B_{{\tau _0}}}
\end{align}
where \textcolor[rgb]{0.00,0.00,0.00}{(\ref{tauknown}) is employed and $\eta$ is defined in (\ref{eta})}.
Similarly, any unbiased joint estimate of time delay, Doppler shift and signal samples satisfies \cite{kay}
\begin{align}\label{fdef}
{\mathop{\rm var}} ({\hat{f}_0}) \ge = JCR{B_{{f_0},{s}}}% MathType!End!2!1!
\end{align}
where \textcolor[rgb]{0.00,0.00,0.00}{\cite{petersen2008matrix}}
\begin{align}\label{fands}
&JCR{B_{{f_0},{s}}}= \left[I(\bm\theta )^{ - 1}\right]_{2,2} \nonumber\\
&= \left[(\bm{A} - \bm{B}{\bm{C}^{ - 1}}{\bm{B}^T})^{ - 1} \right]_{2,2}= \left[ \frac{adj(\bm{A} - \bm{B}{\bm{C}^{ - 1}}{\bm{B}^T})}{det(\bm{A} - \bm{B}{\bm{C}^{ - 1}}{\bm{B}^T})} \right]_{2,2}\nonumber\\
&= {\frac{{LP}}{{L + P}}\frac{2}{{\sigma _w^2}}{\sum\limits_{n = 0}^{M - 1} {{{\left| {\frac{{\partial s(t)}}{{\partial t}}} \right|}^2}\biggr{|}_{t=n \Delta}} }} \cdot\nonumber\\
&\frac{1}{{\frac{{LP}}{{L + P}}{\frac{{8{\pi ^2}}}{{\sigma _w^2}}\sum\limits_{n = 0}^{M - 1} {{{(t + {\tau _0})}^2}{{\left| {s(t)} \right|}^2}\biggr{|}_{t=n \Delta}} }}{{\frac{{LP}}{{L + P}}\frac{2}{{\sigma _w^2}}{\sum\limits_{n = 0}^{M - 1} {{{\left| {\frac{{\partial s(t)}}{{\partial t}}} \right|}^2}}\biggr{|}_{t=n \Delta} }}}-{\left( {\frac{{4\pi \eta }}{{\sigma _w^2}}\frac{{LP}}{{L + P}}} \right)}^2}\nonumber\\
&= \frac{{L + P}}{{LP}} \cdot \frac{{\frac{{\sigma _w^2}}{{8{\pi ^2}}}\sum\limits_{n = 0}^{M - 1} {{{\left| {\frac{{\partial s(t)}}{{\partial t}}} \right|}^2}\biggr{|}_{t=n \Delta}} }}{{\sum\limits_{n = 0}^{M - 1} {{{\left| {\frac{{\partial s(t)}}{{\partial t}}} \right|}^2}\biggr{|}_{t=n \Delta}} \sum\limits_{n = 0}^{M - 1} {{{(t + {\tau _0})}^2}{{\left| {s(t)} \right|}^2}\biggr{|}_{t=n \Delta}}  - {\eta ^2}}}\nonumber\\
&=\frac{L + P}{LP}JCR{B_{{f _0}}}
\end{align}
where \textcolor[rgb]{0.00,0.00,0.00}{(\ref{fknown}) is employed and $\eta$ is defined in (\ref{eta})}.
\textcolor[rgb]{0.00,0.00,0.00}{If we estimate $\tau_0$ and $f_0$ separately for unknown signals, \textcolor[rgb]{0.00,0.00,0.00}{whose CRB we denote as $CR{B_{{\tau_0},{s}}}$ and $CR{B_{{f_0},{s}}}$ respectively}, then (\ref{tauands}) and (\ref{fands}) still hold. Thus,
\begin{align}\label{knastau}
CR{B_{{\tau_0},{s}}}=\frac{L+P}{LP}CR{B_{{\tau _0}}},
\end{align}
\textcolor[rgb]{0.00,0.00,0.00}{with
\begin{align}
CR{B_{{\tau _0}}} = \frac{{\sigma _w^2}}{{2\sum\limits_{n = 0}^{M - 1} {{{\left| {\frac{{\partial s(t)}}{{\partial t}}} \right|}^2}\biggr{|}_{t=n \Delta}} }},
\end{align}}
and
\begin{align}\label{knastau1}
CR{B_{{f_0},{s}}}=\frac{L+P}{LP}CR{B_{{f _0}}},
\end{align}}
\textcolor[rgb]{0.00,0.00,0.00}{with
\begin{align}
CR{B_{{f_0}}} = \frac{{\sigma _w^2}}{{8{\pi ^2}\sum\limits_{n = 0}^{M - 1} {{(t+{\tau _0})^2}{{\left| {s(t)} \right|}^2}\biggr{|}_{t=n \Delta}} }}.
\end{align}}

For $L=0$ and any \textcolor[rgb]{0.00,0.00,0.00}{finite} $P \geq 1$, (\ref{tauands}) and (\ref{fands})
imply there is no unbiased \textcolor[rgb]{0.00,0.00,0.00}{joint} estimator of the time delay $ \tau_0 $ and Doppler shift $ f_0 $. Similarly, if $P=0$ \textcolor[rgb]{0.00,0.00,0.00}{with any finite $L\geq 1$}, (\ref{tauands}) and (\ref{fands})
imply there is no unbiased \textcolor[rgb]{0.00,0.00,0.00}{joint} estimator of the time delay $ \tau_0 $ and Doppler shift $ f_0 $.
This is reasonable.  In fact, it makes sense that we need to see at least one look at the \textcolor[rgb]{0.00,0.00,0.00}{delayed Doppler} shifted and \textcolor[rgb]{0.00,0.00,0.00}{undelayed nonDoppler} shifted signals to provide an accurate joint estimation.  It is clear that $ JCRB_{\tau_0,s} $ and $ JCRB_{f_0,s} $ are decreasing in either $L$ or $P$. From the symmetry of (\ref{tauands}) and (\ref{fands}), the effect of increasing either $L$ or $P$ is exactly the same, as we might expect\textcolor[rgb]{0.00,0.00,0.00}{\footnote{See Section \uppercase\expandafter{\romannumeral 6} for different direct and reflected path channels.}}.

If we want to compare to the known signal case, (\ref{tauknown}) and (\ref{fknown}), we should
recall we just had one look at the time delayed and Doppler shifted version of the signal in that case so it seems $P=1$ should be considered to be fair.
In this case (\ref{tauands}) and (\ref{fands}) imply that $ JCRB_{\tau_0,s}$ and $JCRB_{f_0,s}$  are generally larger than $ JCRB_{\tau_0}$ and $JCRB_{f_0}$ for finite $L$, respectively.
 In fact the factor
$ \left( \frac{L+ P}{L P}  \right)|_{P=1} $ captures the exact increase in a beautiful and simple expression.
 This is interesting since for either $ L = 1 $ or $ P = 1 $, $ JCRB_{\tau_0,s} $ and $JCRB_{f_0,s}$ approach $ JCRB_{\tau_0} $ and $JCRB_{f_0}$ respectively as the other variable (number of looks) approaches infinity.  If $ L > 1 $ or $ P > 1 $, then $ JCRB_{\tau_0,s} $ and $JCRB_{f_0,s}$ approach a value smaller than
$ JCRB_{\tau_0} $ and $JCRB_{f_0}$ respectively as the other variable approaches infinity.   The reason is that $ JCRB_{\tau_0} $ and $JCRB_{f_0}$ are calculated with
 only one look.  If you generalize the model in (\ref{obs}) to $R$ looks, then the JCRB can easily be seen to be
 $ \frac{JCRB}{R} $. Thus the stated limits of $\frac{JCRB_{\tau_0}}{P} $ and $\frac{JCRB_{f_0}}{P} $ as $ L $ increases towards infinity or
 $ \frac{JCRB_{\tau_0}}{L} $ and $ \frac{JCRB_{f_0}}{L} $ as $ P $ increases towards infinity should be expected.  On the other hand, if $L=P$ then by
    making their common value sufficiently large, we can make $ JCRB_{\tau_0,s} $ and $ JCRB_{f_0,s} $ as close to zero as we like.  This seems reasonable since in this case we can perfectly categorize both the \textcolor[rgb]{0.00,0.00,0.00}{undelayed non-Doppler} shifted and the \textcolor[rgb]{0.00,0.00,0.00}{delayed Doppler} shifted signal.
\section{Known signal structure}\label{knownstruc}

Suppose $s(t)$ is a communication signal with known structure and unknown parameters containing information. For example, assume a pulse amplitude modulation signal with unknown complex \textcolor[rgb]{0.00,0.00,0.00}{pulse} amplitudes $b_q$, $q=1,...,Q$ such that $ s(n\Delta)=0$ for $n<0$ and $n>M-1$ and for \textcolor[rgb]{0.00,0.00,0.00}{$n_0 \leq  n \leq n_0 + M - 1$}
\begin{eqnarray}
s(n\Delta-\tau_0) = \sum_{q=1}^{Q} b_q g( n\Delta-\tau_0 - (q-1)T_p )  \label{sigwbits}
\end{eqnarray}
where we assume we know the pulse shape $g(t)$. Further the total support of the signal is still $M$ samples where $M\Delta = QT_P$. In this case, we need to estimate the \textcolor[rgb]{0.00,0.00,0.00}{pulse} amplitudes instead of the signal samples. Thus the parameter to estimate becomes
$\bm\theta = (\tau_0, f_0, b_{1R}, b_{1I}, b_{2R}, b_{2I} ,..., b_{QR}, b_{QI})^T$ where $b_{qR}$ and $b_{qI}$ are the real and imagery parts of the complex \textcolor[rgb]{0.00,0.00,0.00}{pulse} amplitudes $b_q, q=1,...,Q$. It should be noted that $Q$, the number of pulses, is smaller than the number of signal samples used if there is more than one sample per pulse, reasonable for unknown signals. Thus the estimation performance when only estimating the \textcolor[rgb]{0.00,0.00,0.00}{pulse} amplitudes should be more favorable \textcolor[rgb]{0.00,0.00,0.00}{since we estimate fewer parameters} as we show next. While we assume the pulse amplitudes \textcolor[rgb]{0.00,0.00,0.00}{can take on any complex value}, the analysis gives a good approximation if the real and imaginary parts of pulse amplitudes are discrete with many levels.

Define
\begin{eqnarray}\label{hq}
h(q) = \sum_{n=0}^{M-1} \left( \left( \frac{d}{d t} s(t) \right) \biggr{|}_{t=n \Delta} g( n\Delta - (q-1) T_p) \right) \nonumber,
\end{eqnarray}
\begin{eqnarray}\label{uq}
u(q) = \sum_{n=0}^{M-1} \left( \left( s(t) \right)  (t+\tau_0)\biggr{|}_{t=n \Delta}g( n\Delta - (q-1) T_p) \right) \nonumber,
\end{eqnarray}
and
\begin{eqnarray}\label{cq}
c(q,q') =
\sum_{n=0}^{M-1} \left( g( n\Delta - (q-1) T_p) g( n\Delta - (q'-1) T_p) \right).
%\nonumber \\
\end{eqnarray}
The FIM for estimating $\bm\theta$ in this case is
\begin{align}\label{FIMM}
I(\bm\theta ) = \left[ {\begin{array}{*{20}{c}}
{{\bm{A}}}&{{\bm{B}' }}\\
{\bm{B}'^T}&{{\bm{C}' }}
\end{array}} \right]
\end{align}
where the expressions of the specific elements in the ${2 \times 2}$ symmetric matrix ${\bm{A}}$ are given in (\ref{A11})--(\ref{A12}). The specific entries in the ${2 \times 2Q}$ matrix ${\bm{B}'}$ \textcolor[rgb]{0.00,0.00,0.00}{for $q=1,...,Q$} are
\begin{equation}
\bm{B}_{1,j}'=
\begin{cases}
-\mathbb E[\frac{{{\partial ^2}\ln p(\bm{x};\bm\theta )}}{{\partial {b_R(n \Delta)}\partial {\tau_0}}}]=- \frac{{2P}}{{\sigma _w^2}}h_{R}(q)&\mbox{if $ j=2q-1$}\\
-\mathbb E[\frac{{{\partial ^2}\ln p(\bm{x};\bm\theta )}}{{\partial {b_I(n \Delta)}\partial {\tau_0}}}] =- \frac{{2P}}{{\sigma _w^2}}h_{I}(q)&\mbox{if $j=2q$},
\end{cases}
\end{equation}
and
\begin{equation}
\bm{B}_{2,j}'=
\begin{cases}
-\mathbb E[\frac{{{\partial ^2}\ln p(\bm{x};\bm\theta )}}{{\partial {b_R(n \Delta)}\partial {f_0}}}]= - \frac{{4\pi P}}{{\sigma _w^2}}u_{I}(q) &\mbox{if $ j=2q-1$}\\
-\mathbb E[\frac{{{\partial ^2}\ln p(\bm{x};\bm\theta )}}{{\partial {b_I(n \Delta)}\partial {f_0}}}]=  \frac{{4\pi P}}{{\sigma _w^2}}u_{R}(q) &\mbox{if $j=2q$},
\end{cases}
\end{equation}
where $h_{R}(q)$, $h_{I}(q)$ are the real and imaginary parts of $h(q)$ defined in (\ref{hq}),  and $u_{R}(q)$, $u_{I}(q)$ are the real and imaginary parts of $u(q)$ defined in (\ref{uq}).
The entries in the ${2Q \times 2Q}$ matrix ${\bm{C}'}$ for \textcolor[rgb]{0.00,0.00,0.00}{$q, q' = 1,\ldots, Q$} are
\begin{equation}
{\bm{C}}_{2q-1,2q'-1}' = -\mathbb E[\frac{{{\partial ^2}\ln p(x;\theta )}}{{\partial {b_{qR}}\partial {b_{q'R}}}}]= \frac{{2(P + L)}}{{\sigma _w^2}}c(q,q'),
 \end{equation}
 \begin{equation}
{\bm{C}}_{2q,2q'}' = -\mathbb E[\frac{{{\partial ^2}\ln p(x;\theta )}}{{\partial {b_{qI}}\partial {b_{q'I}}}}]= \frac{{2(P + L)}}{{\sigma _w^2}}c(q,q'),
\end{equation}
and
\begin{equation}
{\bm{C}}_{2q-1,2q'}' = -\mathbb E[\frac{{{\partial ^2}\ln p(x;\theta )}}{{\partial {b_{qR}}\partial {b_{q'I}}}}] = 0.
\end{equation}
We can simplify the expressions considerably if $g(n\Delta)$ is nonzero only over the duration of $0< n\Delta < T_P =n_{p}\Delta$, which is often a reasonable approximation. In this case, the FIM is
\begin{align}\label{lalalFIM}
I(\bm\theta ) = \left[ {\begin{array}{*{20}{c}}
{{\bm{A}}}&{{\bm{B}'' }}\\
{\bm{B}''^T}&{{\bm{C}'' }}
\end{array}} \right]
\end{align}
where the elements in the ${2 \times 2}$ symmetric matrix ${{\bm{A}}}$ are given in (\ref{A11})--(\ref{A12}). The entries in the ${2 \times 2Q}$ matrix ${{\bm{B}'' }}$ and those in the ${2Q \times 2Q}$ \textcolor[rgb]{0.00,0.00,0.00}{diagonal} matrix ${{\bm{C}'' }}$ are
\begin{small}
\begin{align}\label{{B}''}
&{{\bm{B}'' }}=\nonumber\\
 &\left[ {\begin{array}{*{20}{c}}
{ - 2P\frac{{\rho b_{1R}}}{{\sigma _w^2}}}&{ - 2P\frac{{\rho b_{1I}}}{{\sigma _w^2}}}&{ - 2P\frac{{\rho b_{2R}}}{{\sigma _w^2}}} &\cdots &{ - 2P\frac{{\rho b_{QI}}}{{\sigma _w^2}}}\\
{ - 4\pi P\frac{\gamma_{1}{b_{1I}}}{{\sigma _w^2}}}&{ 4\pi P\frac{\gamma_{1}{b_{1R}}}{{\sigma _w^2}}}&{ - 4\pi P\frac{\gamma_{2}{b_{2I}}}{{\sigma _w^2}}} & \cdots &{  4\pi P\frac{\gamma_{Q}{b_{QR}}}{{\sigma _w^2}}}\\
\end{array}} \right],
\end{align}
\end{small}
and
\begin{eqnarray}\label{{C}''}
\bm{C}_{j,j} = {\frac{{(2L + 2P)}E_g}{{\sigma _w^2}}} \qquad &\mbox{if $j=1,2,...,2Q$}
\end{eqnarray}
%\begin{eqnarray}\label{{C}''}
%{{\bm{C}'' }}= \left[ {\begin{array}{*{20}{c}}
%{\frac{{(2L + 2P)}E_g}{{\sigma _w^2}}}&0& 0&\cdots &0\\
%0&{\frac{{(2L + 2P)E_g}}{{\sigma _w^2}}}& 0&\cdots  & 0 \\
%\vdots & \vdots &\vdots & \ddots &\vdots\\
%0& 0 &0&\cdots&{\frac{{(2L + 2P)E_g}}{{\sigma _w^2}}}
%\end{array}} \right]
%\end{eqnarray}
with
\begin{eqnarray}\label{rho}
\rho = \sum_{n=0}^{n_p} \left( \left( \frac{d}{d t} g(t) \right) \biggr{|}_{t=n \Delta} g( n\Delta ) \right), \label{rho}
\end{eqnarray}
\begin{eqnarray}\label{gamma}
\gamma_{q}  = \sum\limits_{n = 0}^{{n_p}} {(n\Delta  + {\tau _0}+(q-1)T_P){{(g(n\Delta ))}^2}},
\end{eqnarray}
and
\begin{eqnarray}\label{E_g}
E_g = \sum_{n=0}^{n_p} \left( g( n\Delta) \right)^2.
\end{eqnarray}
Similar to (\ref{FIMinverse}), by using the Schur complement relation and (\ref{lalalFIM}), we obtain
\begin{align}\label{knownstrc}
{[I(\bm\theta )^{ - 1}]_{(\{1,2\},\{1,2\})}}= {{{({{\bm{A}}} - {{\bm{B}'' }}{({{\bm{C}'' }})^{ - 1}}{\bm{B}''^T})}^{ - 1}}}
\end{align}
where ${[I(\bm\theta )^{ - 1}]_{(\{1,2\},\{1,2\})}}$ denotes the sub-matrix of $I(\bm\theta )^{ - 1}$ which consists of the elements located in the first two rows and the first two columns and \textcolor[rgb]{0.00,0.00,0.00}{we \textcolor[rgb]{0.00,0.00,0.00}{define} a $2\times 2$ matrix ${\bm{V}}={{{({{\bm{A}}} - {{\bm{B}'' }}{({{\bm{C}'' }})^{ - 1}}{\bm{B}''^T})}}}$} where
\begin{small}
\begin{align}\label{f11}
{\bm{V}}_{1,1}=\frac{{2P}}{{\sigma _w^2}}\left({{ \sum\limits_{q = 1}^Q {{{\left| {{b_q}} \right|}^2}} }}\sum\limits_{n = 0}^{{n_p}} {{{(\frac{{dg(t)}}{{dt}})}^2}} \biggr{|}_{t=n \Delta} - \frac{{{P}}}{{(L + P)}}\frac{{\rho ^2}}{{E_g}}\sum\limits_{q = 1}^Q {{{\left| {{b_q}} \right|}^2}}\right),
\end{align}
\end{small}
\vspace{-1em}
\begin{small}
\begin{align}\label{f22}
{\bm{V}}_{2,2}&=\frac{{8{\pi ^2}P}}{{\sigma _w^2}}\left( \sum\limits_{q = 1}^Q {( {\sum\limits_{n = 0}^{{n_p}} {{{(t + {\tau _0} + (q - 1){T_p})}^2} \cdot {{(g(t))}^2}} } )\biggr{|}_{t=n \Delta}{{\left| {{b_q}} \right|}^2}}\right.\nonumber\\
&\left. - \frac{P}{L+P}\frac{1}{E_g}\sum\limits_{q = 1}^Q {{{\gamma_q ^2}{\left| {{b_q}} \right|}^2}} \right),
\end{align}
\end{small}
\vspace{-1em}
and
\begin{small}
\begin{align}\label{ccc}
{\bm{V}}_{2,1}&={\bm{V}}_{1,2}= {{\bm{A}}_{1,2}} - \sum\limits_{j = 1}^{2Q} {{{\bm{B}}_{1,j}''}{{\left[ {{{\bm{C}}''^{ - 1}}} \right]}_{j,j}}{\bm{B}}_{j,2}^{''T}}\nonumber \\
&=\frac{{4\pi P}}{{\sigma _w^2}}\eta  - \frac{{\sigma _w^2}}{{(2L + 2P){E_g}}}\sum\limits_{q = 1}^Q {\left( {8\pi {P^2}\frac{{\rho {b_{qR}}}}{{\sigma _w^2}}\frac{{{\gamma _q}{b_{qI}}}}{{\sigma _w^2}} - 8\pi {P^2}\frac{{\rho {b_{qI}}}}{{\sigma _w^2}}\frac{{{\gamma _q}{b_{qR}}}}{{\sigma _w^2}}} \right)}\nonumber \\
&=\frac{{4\pi P}}{{\sigma _w^2}}\eta \nonumber \\
&= \frac{{4\pi P}}{{\sigma _w^2}}\sum\limits_{n = 0}^{M - 1} {(t + {\tau _0})\left[ {\sum\limits_{q = 1}^Q {{b_{qI}}g(t - (q - 1){T_p})}  \cdot \sum\limits_{q = 1}^Q {{b_{qR}}\frac{{\partial g(t - (q - 1){T_p})}}{{\partial t}}}  }\right.}\nonumber\\
&{\left.{- \sum\limits_{q = 1}^Q {{b_{qR}}g(t - (q - 1){T_p})}  \cdot \sum\limits_{q = 1}^Q {{b_{qI}}\frac{{\partial g(t - (q - 1){T_p})}}{{\partial t}}} } \right]} \Biggr{|}_{t=n \Delta} \nonumber \\
&= \frac{{4\pi P}}{{\sigma _w^2}}\sum\limits_{n = 0}^{M - 1} {(t + {\tau _0})\left[ {\sum\limits_{q = 1}^Q {{b_{qI}}{b_{qR}}g(t - (q - 1){T_p})}  \cdot \frac{{\partial g(t - (q - 1){T_p})}}{{\partial t}} }\right.}\nonumber\\
&{\left.{ - \sum\limits_{q = 1}^Q {{b_{qR}}{b_{qI}}g(t - (q - 1){T_p})}  \cdot \frac{{\partial g(t - (q - 1){T_p})}}{{\partial t}}} \right]} \Biggr{|}_{t=n \Delta}\nonumber\\
&= 0
\end{align}
\end{small}where we have used the results in (\ref{A12}), (\ref{{B}''}) and (\ref{{C}''}) to \textcolor[rgb]{0.00,0.00,0.00}{obtain} the second line. Since the second term in the second line is zero, we \textcolor[rgb]{0.00,0.00,0.00}{obtain} the third line. We have used the known signal structure to obtain the fourth line and we used the assumption that $g(n\Delta)$ is nonzero only over the duration of $0< n\Delta< T_P =n_{p}\Delta$ to obtain the fifth line.

Thus when performing the joint estimation of time delay, Doppler shift and the complex pulse amplitudes
\begin{align}\label{timeamplitude}
{\mathop{\rm var}} ({\hat{\tau}_0}) \ge  JCR{B_{{\tau_0},b}}% MathType!End!2!1!
\end{align}
where $JCR{B_{{\tau_0},b}}$ is calculated as
\begin{align}\label{CRBtaub}
&JCR{B_{{\tau _0},b}} = [I(\bm\theta )^{ - 1}]_{1,1}\nonumber\\
&= \frac{{\sigma _w^2}}{{2P}}{\left( { {\sum\limits_{q = 1}^Q {{{\left| {{b_q}} \right|}^2}} } }\sum\limits_{n = 0}^{{n_p}} {{{(\frac{dg(t)}{{dt}})^2}\biggr{|}_{t=n \Delta}}}  - \frac{P}{{L + P}}\frac{{{\rho ^2}}}{{{E_g}}}\sum\limits_{q = 1}^Q {{{\left| {{b_q}} \right|}^2}} \right)^{ - 1}}.
\end{align}
Similar to (\ref{timeamplitude}), any unbiased estimate of time delay, Doppler shift and the complex pulse amplitudes together satisfies
\begin{align}\label{Doppleramplitude}
{\mathop{\rm var}} ({\hat{f}_0}) \ge JCR{B_{{f_0},b}} % MathType!End!2!1!
\end{align}
where $JCR{B_{{f_0},b}}$ is derived as
\begin{align}\label{CRBfb}
&JCR{B_{{f_0},b}} = [I(\bm\theta )^{ - 1}]_{2,2}\nonumber\\
&= \frac{{\sigma _w^2}}{{8{\pi ^2}P}}\left({{\sum\limits_{q = 1}^Q {( {\sum\limits_{n = 0}^{{n_p}} {{{(t + {\tau _0} + (q - 1){T_p})}^2} \cdot {{(g(t))}^2}} } )\biggr{|}_{t=n \Delta}{{\left| {{b_q}} \right|}^2}}}}\right.\nonumber\\
&\left.{{  - \frac{P}{{L + P}}\frac{1}{{{E_g}}}\sum\limits_{q = 1}^Q {\gamma _q^2{{\left| {{b_q}} \right|}^2}} }}\right)^{ - 1}.
\end{align}
\textcolor[rgb]{0.00,0.00,0.00}{For the case where $g(n\Delta)$ is nonzero only over the duration of $0< n\Delta < T_P =n_{p}\Delta$}, the JCRB of the time delay and Doppler shift estimation with known signals in (\ref{tauknown}) becomes
\begin{align}\label{CRBtaus}
JCR{B_{{\tau _0}}} = \frac{{\sigma _w^2}}{{2{\sum\limits_{q = 1}^Q {{{\left| {{b_q}} \right|}^2}} } \sum\limits_{n = 0}^{{n_p}} {{{(\frac{{dg(t)}}{{dt}})}^2}\biggr{|}_{t=n \Delta}}}}.
\end{align}
Similar to (\ref{CRBtaus}), the JCRB of the time delay and Doppler shift estimation with known signals satisfies
\begin{align}
JCR{B_{{f_0}}} = \frac{{\sigma _w^2}}{{8{\pi ^2}\sum\limits_{q = 1}^Q {\left( {\sum\limits_{n = 0}^{{n_p}} {{{(t + {\tau _0} + (q - 1){T_p})}^2} \cdot {{(g(t))}^2}} } \right)\biggr{|}_{t=n \Delta}{{\left| {{b_q}} \right|}^2}} }}.
\end{align}

Provided \textcolor[rgb]{0.00,0.00,0.00}{\textbf{$\bf g(\textit{t}) $ is not equal to a scalar multiple of $\bf\frac{d}{d \textit{t}} g(\textit{t}) $ and \textcolor[rgb]{0.00,0.00,0.00}{$\bm{{n_p}\ne0}$}}},  from the Schwartz inequality, (\ref{rho}) and (\ref{E_g}), we find
\begin{eqnarray}\label{rhorho}
 \rho^2 <   E_g
 \sum_{n=0}^{n_p} \left( \left( \frac{d}{d t} g(t) \right)^2 \biggr{|}_{t=n \Delta} \right).
\end{eqnarray}
For $P>0$, $L>0$, multiplying both sides by $ - \frac{P}{{L + P}} \cdot \frac{1}{{{E_g}}}\sum\limits_{q = 1}^Q {{{\left| {{b_q}} \right|}^2}} $ implies
\begin{eqnarray}\label{rrrho}
- \frac{P}{{L + P}} \cdot \frac{{{\rho ^2}}}{{{E_g}}}\sum\limits_{q = 1}^Q {{{\left| {{b_q}} \right|}^2}}  >  - \frac{P}{{L + P}} \cdot \sum\limits_{q = 1}^Q {{{\left| {{b_q}} \right|}^2}}\sum\limits_{n = 0}^{{n_p}} {{{(\frac{{dg(t)}}{{dt}})^2}\biggr{|}_{t=n \Delta}}} .
\end{eqnarray}
%Assuming all real parts of the amplitudes have the same sign, so do all imaginary parts implies
%\begin{eqnarray}\label{sssho}
%\sum\limits_{q = 1}^Q {{{\left| {{b_q}} \right|}^2}}  < {\left| {\sum\limits_{q = 1}^Q {{b_q}} } \right|^2}
%\end{eqnarray}
%Combining (\ref{rrrho}) with (\ref{sssho}), it implies
%\begin{eqnarray}
% - \frac{P}{{L + P}} \cdot \frac{{{\rho ^2}}}{{{E_g}}}\sum\limits_{q = 1}^Q {{{\left| {{b_q}} \right|}^2}}  >  - \frac{P}{{L + P}} \cdot \sum\limits_{n = 0}^{{n_p}} {{{(\frac{{dg(t)}}{{dt}})}^2}} {\left| {\sum\limits_{q = 1}^Q {{b_q}} } \right|^2}
%\end{eqnarray}
Adding ${ {\sum\limits_{q = 1}^Q {{{\left| {{b_q}} \right|}^2}} } }\sum\limits_{n = 0}^{{n_p}} {{{(\frac{dg(t)}{{dt}})^2}\biggr{|}_{t=n \Delta}}}$ to both sides yields,
\begin{eqnarray}\label{vvvho}
&{ {\sum\limits_{q = 1}^Q {{{\left| {{b_q}} \right|}^2}} } }\sum\limits_{n = 0}^{{n_p}} {{{(\frac{dg(t)}{{dt}})^2}\biggr{|}_{t=n \Delta}}}  - \frac{P}{{L + P}} \cdot \frac{{{\rho ^2}}}{{{E_g}}}\sum\limits_{q = 1}^Q {{{\left| {{b_q}} \right|}^2}}  >\nonumber\\
&{ {\sum\limits_{q = 1}^Q {{{\left| {{b_q}} \right|}^2}} } }\sum\limits_{n = 0}^{{n_p}} {{{(\frac{dg(t)}{{dt}})^2}\biggr{|}_{t=n \Delta}}}  - \frac{P}{{L + P}} \cdot { {\sum\limits_{q = 1}^Q {{{\left| {{b_q}} \right|}^2}} } }\sum\limits_{n = 0}^{{n_p}} {{{(\frac{dg(t)}{{dt}})^2}\biggr{|}_{t=n \Delta}}}.
\end{eqnarray}
Calculating the reciprocal of both sides of (\ref{vvvho}) and multiplying both sides of them by $\frac{{\sigma _w^2}}{{2P}}$ implies
\begin{eqnarray}\label{sssh}
&\frac{{\sigma _w^2}}{{2P}}\cdot{\left( { {\sum\limits_{q = 1}^Q {{{\left| {{b_q}} \right|}^2}} } }\sum\limits_{n = 0}^{{n_p}} {{{(\frac{dg(t)}{{dt}})^2}\biggr{|}_{t=n \Delta}}}  - \frac{P}{{L + P}} \cdot \frac{{{\rho ^2}}}{{{E_g}}}\sum\limits_{q = 1}^Q {{{\left| {{b_q}} \right|}^2}} \right)^{ - 1}} <\frac{{\sigma _w^2}}{{2P}}\cdot \nonumber\\
&{\left( { {\sum\limits_{q = 1}^Q {{{\left| {{b_q}} \right|}^2}} } }\sum\limits_{n = 0}^{{n_p}} {{{(\frac{dg(t)}{{dt}})^2}\biggr{|}_{t=n \Delta}}}  - \frac{P}{{L + P}} \cdot { {\sum\limits_{q = 1}^Q {{{\left| {{b_q}} \right|}^2}} } }\sum\limits_{n = 0}^{{n_p}} {{{(\frac{dg(t)}{{dt}})^2}\biggr{|}_{t=n \Delta}}} \right)^{ - 1}}.
\end{eqnarray}
According to (\ref{CRBtaub}), the left side of (\ref{sssh}) is $JCRB_{\tau_0,b}$. \textcolor[rgb]{0.00,0.00,0.00}{Using} (\ref{CRBtaus}), the right side of (\ref{sssh}) is $\frac{L+P}{LP}JCR{B_{{\tau _0}}}$.
This implies \textcolor[rgb]{0.00,0.00,0.00}{(given bold above (\ref{rhorho}))} $ JCRB_{\tau_0,b} \textcolor[rgb]{0.00,0.00,0.00}{<} JCRB_{\tau_0,s} $ from previous results in (\ref{tauands}). Similarly, we can show $ JCRB_{f_0,b} \textcolor[rgb]{0.00,0.00,0.00}{<} JCRB_{f_0,s} $. Thus, the known signal structure will help to improve the estimation performance compared with \textcolor[rgb]{0.00,0.00,0.00}{that from} totally unknown signals.

Noting that $ {\cal P} = \frac{ 2P^2 \rho^2\sum\nolimits_{q = 1}^Q {{{\left| {{b_q}} \right|}^2}}  }{ ( L + P ) E_g } \geq 0 $
and that $ JCRB_{\tau_0,b} $ in (\ref{CRBtaub}) is monotonic
increasing in $ {\cal P} $ for everything else constant\textcolor[rgb]{0.00,0.00,0.00}{,} then the
smallest possible $ JCRB_{\tau_0,b} $ occurs when $ {\cal P} = 0 $.  Interestingly, from (\ref{rho}), $ \rho $ can be zero for symmetric pulse waveforms so such waveforms can produce this smallest possible $ JCRB_{\tau_0,b} $.   In the numerical example section, numerical results will show the magnitude of the performance gains for the known signal structure in (\ref{CRBtaub}) and (\ref{CRBfb}) over totally unknown signals.

\textcolor[rgb]{0.00,0.00,0.00}{If we estimate $\tau_0$ and $f_0$ separately instead of jointly, we can show $CRB_{\tau_0,b}<CRB_{\tau_0,s}$ and $CRB_{f_0,b}<CRB_{f_0,s}$}
\textcolor[rgb]{0.00,0.00,0.00}{where we can show $CR{B_{{\tau _0},b}}=JCR{B_{{\tau _0},b}}$ and $CR{B_{{f_0},b}}=JCR{B_{{f_0},b}}$ since we have shown ${\bm{V}}_{2,1}={\bm{V}}_{1,2}=0$ in (\ref{ccc}).}
%\begin{align}
%CR{B_{{\tau _0},b}} = \frac{{\sigma _w^2}}{{2P}}{\left( { {\sum\limits_{q = 1}^Q {{{\left| {{b_q}} \right|}^2}} } }\sum\limits_{n = 0}^{{n_p}} {{{(\frac{dg(t)}{{dt}})^2}\biggr{|}_{t=n \Delta}}}  - \frac{P}{{L + P}}\frac{{{\rho ^2}}}{{{E_g}}}\sum\limits_{q = 1}^Q {{{\left| {{b_q}} \right|}^2}} \right)^{ - 1}},
%\end{align}
%and
%\begin{align}\label{CRBFBsep}
%CR{B_{{f_0},b}} = \frac{{\sigma _w^2}}{{8{\pi ^2}P}}\left({{\sum\limits_{q = 1}^Q {( {\sum\limits_{n = 0}^{{n_p}} {{{(t + {\tau _0} + (q - 1){T_p})}^2} \cdot {{(g(t))}^2}} } )\biggr{|}_{t=n \Delta}{{\left| {{b_q}} \right|}^2}}  - \frac{P}{{L + P}}\frac{1}{{{E_g}}}\sum\limits_{q = 1}^Q {\gamma _q^2{{\left| {{b_q}} \right|}^2}} }}\right)^{ - 1}.
%\end{align}
%It is worth noting that $CR{B_{{\tau _0},b}}=JCR{B_{{\tau _0},b}}$ and $CR{B_{{f_0},b}}=JCR{B_{{f_0},b}}$ which is reasonable since we have shown ${\bm{V}}_{2,1}={\bm{V}}_{1,2}=0$ in (\ref{ccc}) which means the time delay part of joint estimation has no effect on the Doppler shift part of joint estimation and vice versa. This causes that joint estimation of the time delay and Doppler shift will have the same results as we separately estimate them which means $CR{B_{{\tau _0},b}}=JCR{B_{{\tau _0},b}}$ and $CR{B_{{f_0},b}}=JCR{B_{{f_0},b}}$.

% It is worth noting that $c=0$ in (\ref{ccc}) shows that

\begin{figure}[!t]
\centering
\includegraphics[width=3.0in]{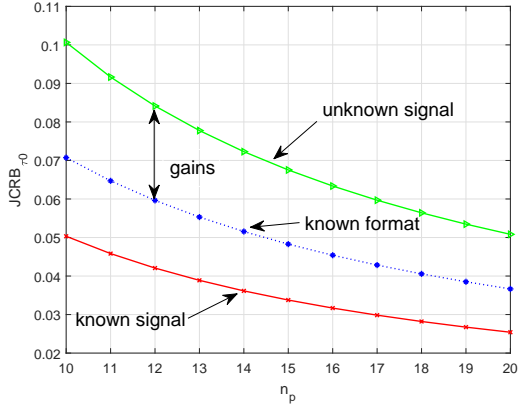}
\caption{$JCRB_{\tau_0}$, $JCRB_{\tau_0,s}$ and $JCRB_{\tau_0,b}$ of the joint estimation for increasing samples $n_p$ when $P=L=1$.}
\label{conv_fig4}
\end{figure}

\begin{figure}[!t]
\centering
\includegraphics[width=3.0in]{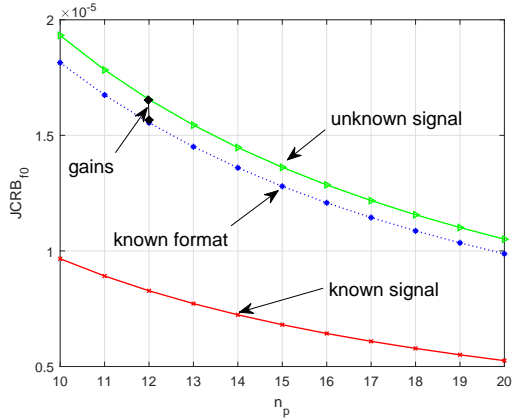}
\caption{$JCRB_{f_0}$, $JCRB_{f_0,s}$ and $JCRB_{f_0,b}$ of the joint estimation for increasing samples $n_p$ when $P=L=1$.}
\label{conv_fig5}
\end{figure}

\textcolor[rgb]{0.00,0.00,0.00}{\section{Numerical Examples}}
Initially consider the case where the transmitted signal $s(t)$ is completely unknown to
the estimator but is described by (\ref{sigwbits}) \textcolor[rgb]{0.00,0.00,0.00}{for $ b_{qR}^2 = b_{qI}^2 = +1$ for} $q = 1,\ldots, Q $ (one of many possible examples).  Assume the case where $g(n\Delta)$ is nonzero only over the duration of $0 < n \Delta < T_p = n_p \Delta $ where $n_p\Delta=T_p=4, n_p=10,11,...,20$. In
particular $ g(n \Delta) = \exp{( -(n\Delta - 4)^2/9 )} $ for $ n = 0,\ldots, n_p $ and zero elsewhere. Let $P=L=1$, $\tau_0 = 0.5$, $\sigma_w^2 = 0.1$, $Q=1$ and $ M = Qn_p $. As a function of $n_p$, the number of samples per pulse, $JCRB_{\tau_0}$, $JCRB_{\tau_0,s}$ and $JCRB_{\tau_0,b}$ of the joint estimation are shown in Fig.~\ref{conv_fig4}. Similar numerical results \textcolor[rgb]{0.00,0.00,0.00}{for} $JCRB_{f_0}$, $JCRB_{f_0,s}$ and $JCRB_{f_0,b}$  are shown in Fig.~\ref{conv_fig5}. These results illustrate the gains of the joint estimation with known signal structure when compared with totally unknown signals. It is worth noting that the case just considered employed a \textcolor[rgb]{0.00,0.00,0.00}{nearly} symmetric pulse communication signal, as might typically be exploited in passive radar,
which produces a small $ \rho $ in (\ref{rho})\textcolor[rgb]{0.00,0.00,0.00}{. On the other hand} $\gamma_{q}$ in (\ref{gamma}) can not be zero so that Fig.~\ref{conv_fig4} and Fig.~\ref{conv_fig5} show \textcolor[rgb]{0.00,0.00,0.00}{ different sized gains between the unknown signals JCRB and the known format JCRB due to the different impact of the second term inside the $()^{-1}$ in (\ref{CRBtaub}) and (\ref{CRBfb}).} For different signals, the results might be different.

%\begin{figure}[htb]
%\centerline{
%\includegraphics[width=0.49\textwidth,height=0.35\textwidth]{figures/CRB_tau_samples.eps}
%%\includegraphics[width=0.45\textwidth]{Triangularly.eps}
%\vspace{-0.10in}
%}
%\caption{$JCRB_{\tau_0}$, $JCRB_{\tau_0,s}$ and $JCRB_{\tau_0,b}$ of the joint estimation for increasing samples $n_p$ when $P=L=1$.}
%\label{conv_fig4}
%\end{figure}
%\begin{figure}[htb]
%\centerline{
%\includegraphics[width=0.49\textwidth,height=0.35\textwidth]{figures/CRB_f_samples.eps}
%%\includegraphics[width=0.45\textwidth]{Triangularly.eps}
%\vspace{-0.10in}
%}
%\caption{$JCRB_{f_0}$, $JCRB_{f_0,s}$ and $JCRB_{f_0,b}$ of the joint estimation for increasing samples $n_p$ when $P=L=1$.}
%\label{conv_fig5}
%\end{figure}

\begin{figure}[!t]
\centering
\includegraphics[width=3.0in]{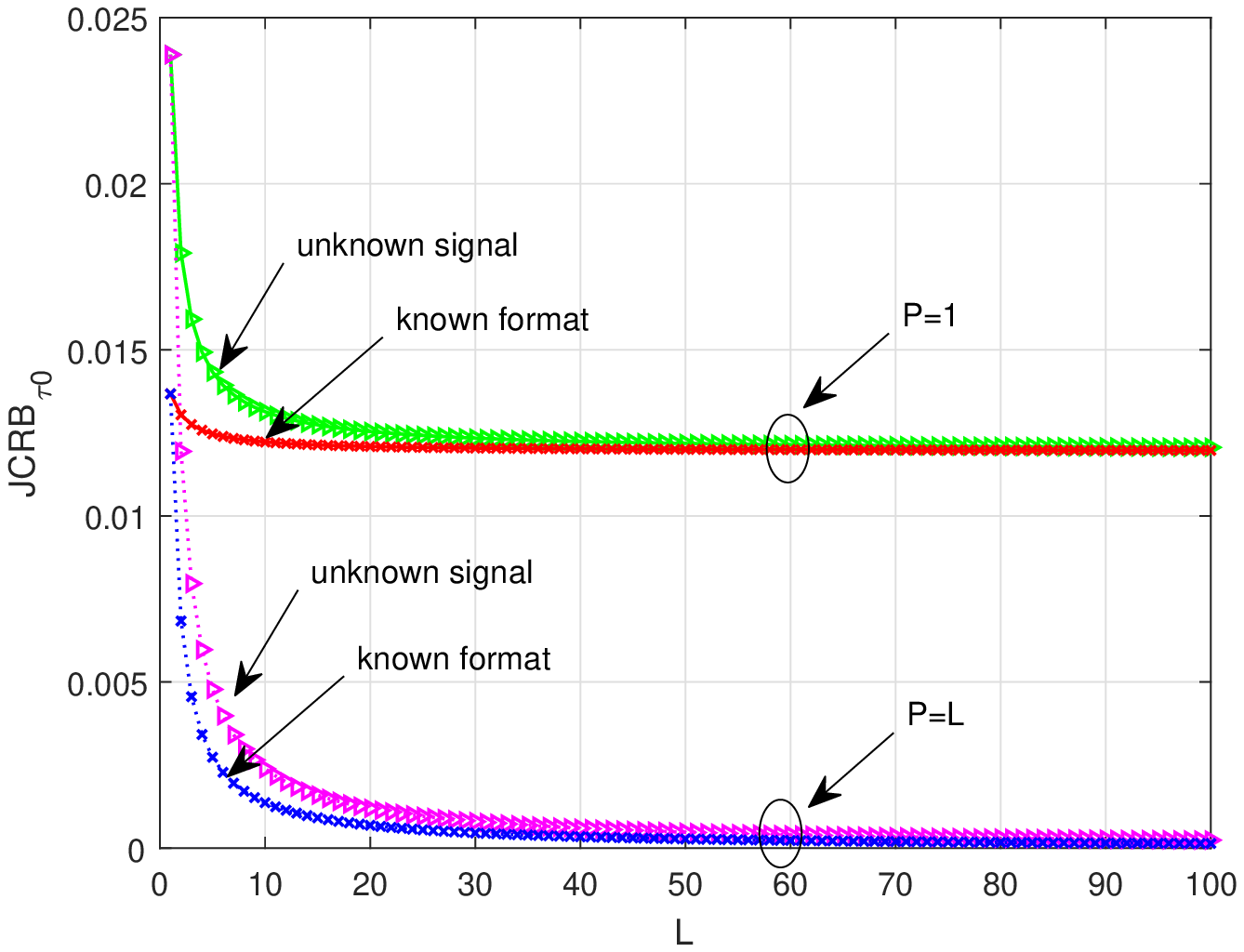}
\caption{As $L$ increases, both the unknown signal $JCRB_{\tau_0,s}$ (unknown form) and the known signal format $JCRB_{\tau_0,b}$ converge to the known signal $JCRB_{\tau_0}$ when $P=1$  and to zero when $P=L$.}
\label{conv_fig}
\end{figure}

\begin{figure}[!t]
\centering
\includegraphics[width=3.0in]{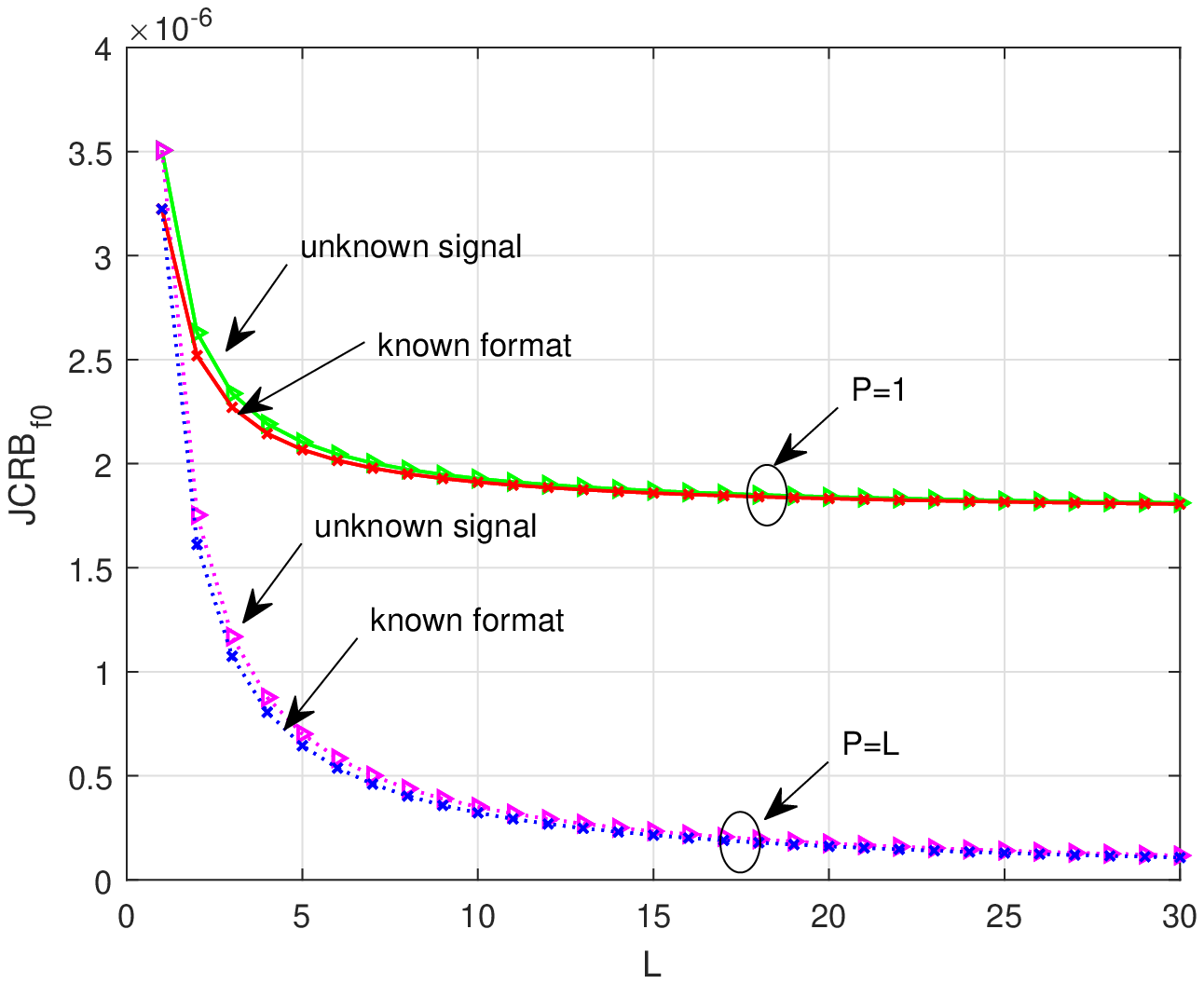}
\caption{As $L$ increases, both the unknown signal $JCRB_{f_0,s}$ (unknown form) and the known signal format $JCRB_{f_0,b}$ converge to the known signal $JCRB_{f_0}$ when $P=1$  and to zero when $P=L$.}
\label{conv_fig2}
\end{figure}

Consider the same signal example but let $\Delta=0.01$, $n_p=500$, $\tau_0 = 0.05$, $f_0=20$, $ Q = 2 $, $ M = Qn_p $  and
$\sigma_w^2 = 1$.  For $P=1$ with expressions in (\ref{tauands}) and (\ref{fands}), we find $ JCRB_{\tau_0} $, $ JCRB_{\tau_0,s} $, $ JCRB_{f_0} $  and $ JCRB_{f_0,s} $ as  shown in Table~\ref{table:TSLATDS}.  This indicates that $ JCRB_{\tau_0,s} $, $ JCRB_{f_0,s} $ approach $ JCRB_{\tau_0} $, $ JCRB_{f_0} $ respectively for large $L$ and $P=1$ as expected. This is also shown in the top curve in Fig.~\ref{conv_fig} and Fig.~\ref{conv_fig2} which is labeled "unknown
signal $P=1$" for the time delay and Doppler shift, respectively.
The results for different $ \sigma_w^2 $ look very similar to the results in Table~\ref{table:TSLATDS}, Fig.~\ref{conv_fig} and Fig.~\ref{conv_fig2}
with the only difference being the $ \sigma_w^2 $ dependence in (\ref{tauknown}) and (\ref{fknown})  which scales the results.
\begin{table}[ht]
\caption{The CRBs with Unknown and Known Signals Results}
 % title of Table
\centering % used for centering table
\begin{tabular}{c c c c c} % centered columns (4 columns)
\hline\hline %inserts double horizontal lines
$L$ & $ JCRB_{\tau_0,s} $ & $JCRB_{\tau_0}$ & $ JCRB_{f_0,s}(10^{-5}) $ & $JCRB_{f_0}(10^{-5})$\\ [0.5ex] % inserts table
%heading
\hline % inserts single horizontal line
1 & 0.0239 & 0.0119 & 0.3506 & 0.1753\\ % inserting body of the table
2 & 0.0179 & 0.0119 &  0.2629 & 0.1753\\
100 & 0.0121 & 0.0119 & 0.1770 & 0.1753\\[1ex] % [1ex] adds vertical space
\hline %inserts single line
\end{tabular}
\label{table:TSLATDS} % is used to refer this table in the text
\end{table}

%\begin{figure}[htb]
%%\vspace{-0.1in}
%\centerline{
%\includegraphics[width=0.49\textwidth,height=0.35\textwidth]{figures/timedelay.eps}
%%\includegraphics[width=0.45\textwidth]{Triangularly.eps}
%\vspace{-0.10in}
%}
%\caption{\textcolor[rgb]{0.00,0.00,0.00}{As $L$ increases, both the unknown signal $JCRB_{\tau_0,s}$ (unknown form) and the known signal format $JCRB_{\tau_0,b}$ converge to the known signal $JCRB_{\tau_0}$ when $P=1$  and to zero when $P=L$.}}
%\label{conv_fig}
%\end{figure}

%\begin{figure}[htb]
%%\vspace{-0.1in}
%\centerline{
%\includegraphics[width=0.49\textwidth,height=0.35\textwidth]{figures/Dopplershift.eps}
%%\includegraphics[width=0.45\textwidth]{Triangularly.eps}
%\vspace{-0.10in}
%}
%\caption{\textcolor[rgb]{0.00,0.00,0.00}{As $L$ increases, both the unknown signal $JCRB_{f_0,s}$ (unknown form) and the known signal format $JCRB_{f_0,b}$ converge to the known signal $JCRB_{f_0}$ when $P=1$  and to zero when $P=L$.}}
%\label{conv_fig2}
%\end{figure}

%\textcolor[rgb]{1.00,0.00,0.00}{For the cases with $P>1$, we see the same trend in Figure~\ref{conv_fig} if we fix $L=1$ and increase $P$ as expected.  }
If we increase both $P$ and $L$ together we can make $ JCRB_{\tau_0,s} $ smaller
than $JCRB_{\tau_0}$ as expected and for large $P=L$, $ JCRB_{\tau_0,s} $ approaches zero as shown in the second lowest curve in Fig.~\ref{conv_fig}
which is labeled "unknown signal $P=L$". It is similar for \textcolor[rgb]{0.00,0.00,0.00}{the} Doppler shift case which is shown in Fig.~\ref{conv_fig2}.

Consider the same parameters as \textcolor[rgb]{0.00,0.00,0.00}{for the} unknown signals case but now assume only the pulse amplitudes $b_q$, the time delay $\tau_0$ and the Doppler shift $f_0$ are unknown as in Section~\ref{knownstruc}.  The second highest curve in Fig.~\ref{conv_fig}, labeled "known format $P=1$" shows the advantages of knowing the signal format (compared to the highest) and estimating the pulse amplitudes rather than the samples.  This implies we estimate a factor of $n_p=500$ fewer unknowns.   The lowest curve in Fig.~\ref{conv_fig}, labeled "known format $P=L$" shows the advantages of knowing the signal format when $P=L$.
The advantage for large $P=L$ is clear if we compare (\ref{tauands}), which decays as $ \frac{2 JCRB_{\tau_0}}{L} $ when
$P=L$, to (\ref{CRBtaub}), which decays as  $ \frac{ JCRB_{\tau_0}}{L} $ with a small $\rho$ for large $ P=L $. \textcolor[rgb]{0.00,0.00,0.00}{Similar} results for $JCRB_{f_0} $ are shown in Fig.~\ref{conv_fig2}.

\textcolor[rgb]{0.00,0.00,0.00}{\section{Extensions}}

After analysis with \textcolor[rgb]{0.00,0.00,0.00}{the \textcolor[rgb]{0.00,0.00,0.00}{basic} models in (\ref{refobs}) and (\ref{changeobs}) for unknown signals and extensions to signals with known structure, further extensions are studied in this section.}
%part to support the utility of the CRB with unknown signals. Specifically, the closed-form of the CRB with unknown signals is derived in the case in which the signals are overlapped together without isolating them from direct path and reflected path returns. The impact of scales of overlap on the CRB with unknown signals is obtained in numerical results. Besides, we consider the CRB with unknown signals in more practical case in which the radar system have correlated noise and multipath effects. The closed-form of the CRB with unknown signals in correlated noise case can be derived by using the methods we discuss in the following. As for the multipath propagation, the specific estimation strategy is provided for whether we know the channel or not. More importantly, the method we provide to joint estimate time delay and Doppler shift with unknown signals in Section \ref{methodwe} is a basic whether for overlapped signals, correlated noise case or multipath propagation.
\textcolor[rgb]{0.00,0.00,0.00}{\subsection{Different Known and Unknown SCNR for the Direct and Reflected Paths}}
First consider the case where the reflected path \textcolor[rgb]{0.00,0.00,0.00}{model} is different from the model in (\ref{changeobs}), so that the reflected path signal is
\begin{eqnarray}\label{Unknowncofreal1}
 x_{r\ell}(n\Delta) &=& a s(n\Delta-\tau_0){e^{j2\pi {f_0} n\Delta }} + \textcolor[rgb]{0.00,0.00,0.00}{w_{r\ell}(n\Delta)}
\end{eqnarray}
\textcolor[rgb]{0.00,0.00,0.00}{for $n = 0,\ldots, N-1$ and $\ell = 1, \ldots, P$}, where $a$ is a \textcolor[rgb]{0.00,0.00,0.00}{known} factor which characterizes the different SCNR \textcolor[rgb]{0.00,0.00,0.00}{of the reflected path when compared to the direct path}. The SCNR is scaled by the factor $a$ if the variance $\sigma_w^2$ of the clutter-plus-noise is exactly the same as we used in (\ref{refobs}) and (\ref{changeobs}). If we repeat the calculations in Section \uppercase\expandafter{\romannumeral 3} and Section \uppercase\expandafter{\romannumeral 4}, the previous results should be modified to
\begin{align}\label{reflecoefftau}
JCR{B_{{\tau_0},\bm{s}}}=\frac{L+a^2P}{LP}JCR{B_{{\tau_0}}}
\end{align}
\textcolor[rgb]{0.00,0.00,0.00}{with a \textcolor[rgb]{0.00,0.00,0.00}{redefined} $JCR{B_{{\tau_0}}}$
\begin{align}
JCR{B_{{\tau_0}}}=\frac{{\sigma _w^2\sum\limits_{n = 0}^{M - 1} {{{(t + {\tau _0})}^2}{{\left| {s(t)} \right|}^2}}\biggr{|}_{t=n \Delta} }}{{2a^2\left( {\sum\limits_{n = 0}^{M - 1} {{{\left| {\frac{{\partial s(t)}}{{\partial t}}} \right|}^2}\biggr{|}_{t=n \Delta}} \sum\limits_{n = 0}^{M - 1} {{{(t + {\tau _0})}^2}{{\left| {s(t)} \right|}^2}\biggr{|}_{t=n \Delta}}  - {\eta ^2}} \right)}}.
\end{align}}
\textcolor[rgb]{0.00,0.00,0.00}{Further,}
\begin{align}\label{reflecoefff}
JCR{B_{{f_0},\bm{s}}}=\frac{L+a^2P}{LP}JCR{B_{{f_0}}}
\end{align}
with a \textcolor[rgb]{0.00,0.00,0.00}{redefined} $JCR{B_{{f_0}}}$
\begin{align}
JCR{B_{{f_0}}}=\frac{{\sigma _w^2\sum\limits_{n = 0}^{M - 1} {{{\left| {\frac{{\partial s(t)}}{{\partial t}}} \right|}^2}}\biggr{|}_{t=n \Delta} }}{{8{\pi ^2}a^2\left( {\sum\limits_{n = 0}^{M - 1} {{{\left| {\frac{{\partial s(t)}}{{\partial t}}} \right|}^2}\biggr{|}_{t=n \Delta}} \sum\limits_{n = 0}^{M - 1} {{{(t + {\tau _0})}^2}{{\left| {s(t)} \right|}^2}\biggr{|}_{t=n \Delta}}  - {\eta ^2}} \right)}}.
\end{align}
Further, we can again show $JCRB_{\tau_0,b}< JCRB_{\tau_0,s}$ and $JCRB_{f_0,b}<JCRB_{f_0,s}$. \textcolor[rgb]{0.00,0.00,0.00}{If $\tau_0$ and $f_0$ \textcolor[rgb]{0.00,0.00,0.00}{are estimated} separately for unknown signals with the known factor $a$, $CRB_{\tau_0,b}< CRB_{\tau_0,s}$ and $CRB_{f_0,b}<CRB_{f_0,s}$ can also be shown and the results in (\ref{knastau}) and (\ref{knastau1}) become
\begin{align}\label{reflecoetau2}
CR{B_{{\tau_0},\bm{s}}}=\frac{L+a^2P}{LP}CR{B_{{\tau _0}}}
\end{align}
\textcolor[rgb]{0.00,0.00,0.00}{with
\begin{align}\label{reflecoetau2CRB}
CR{B_{{\tau_0}}}=\frac{{\sigma _w^2}}{{2a^2\sum\limits_{n = 0}^{M - 1} {{{\left| {\frac{{\partial s(t)}}{{\partial t}}} \right|}^2}\biggr{|}_{t=n \Delta}} }},
\end{align}}
and
\begin{align}\label{reflecoefff2}
CR{B_{{f_0},\bm{s}}}=\frac{L+a^2P}{LP}CR{B_{{f _0}}}
\end{align}
with
\begin{align}\label{reflecoefff2CRB}
CR{B_{{f_0}}}=\frac{{\sigma _w^2}}{{8{\pi ^2}a^2\sum\limits_{n = 0}^{M - 1} {{(t+\tau_0)^2}{{\left| {s(t)} \right|}^2}\biggr{|}_{t=n \Delta}} }}.
\end{align}
This makes sense since the relative importance of the observations from (\ref{refobs}) and (\ref{Unknowncofreal1}) are different due to the different SCNRs. If the factor $a$ is unknown and needs to be estimated with the other unknowns, we still obtain (\ref{reflecoefftau}) (\ref{reflecoefff}) (\ref{reflecoetau2}) (\ref{reflecoefff2}) and show \textcolor[rgb]{0.00,0.00,0.00}{$JCRB_{\tau_0,b}< JCRB_{\tau_0,s}$, $JCRB_{f_0,b}<JCRB_{f_0,s}$, $CRB_{\tau_0,b}< CRB_{\tau_0,s}$ and $CRB_{f_0,b}<CRB_{f_0,s}$} where
\textcolor[rgb]{0.00,0.00,0.00}{\begin{align}\label{aJCRBtaub}
JCR{B_{{\tau _0},b}} = \frac{{\sigma _w^2}}{{2{a^2}P\sum\limits_{q = 1}^Q {{{\left| {{b_q}} \right|}^2}} }}\frac{{{E_g}}}{{{E_g}\sum\limits_{n = 0}^{{n_p}} {{{\left( {\frac{{dg(t)}}{{dt}}} \right)}^2}{\biggr{|}_{t=n\Delta}} - {\rho ^2}} }},
\end{align}
\begin{align}\label{aJCRBfb}
&JCR{B_{{f_0},b}} = \frac{{\sigma _w^2}}{{8{\pi ^2}a^2P}}\left({{\sum\limits_{q = 1}^Q {( {\sum\limits_{n = 0}^{{n_p}} {{{(t + {\tau _0} + (q - 1){T_p})}^2} \cdot {{(g(t))}^2}} } )\biggr{|}_{t=n \Delta}{{\left| {{b_q}} \right|}^2}} }}\right.\nonumber\\
&\left.{{- \frac{a^2P}{{L + a^2P}}\frac{1}{{{E_g}}}\sum\limits_{q = 1}^Q {\gamma _q^2{{\left| {{b_q}} \right|}^2}} }}\right)^{ - 1},
\end{align}
\begin{align}\label{aCRBtaub}
CR{B_{{\tau _0},b}} = \frac{{\sigma _w^2}}{{2{a^2}P\sum\limits_{q = 1}^Q {{{\left| {{b_q}} \right|}^2}} }}\frac{{{E_g}}}{{{E_g}\sum\limits_{n = 0}^{{n_p}} {{{\left( {\frac{{dg(t)}}{{dt}}} \right)}^2}{\biggr{|}_{t=n\Delta}} - {\rho ^2}} }},
\end{align}
and
\begin{align}\label{aCRBfb}
&CR{B_{{f_0},b}} = \frac{{\sigma _w^2}}{{8{\pi ^2}a^2P}}\left({{\sum\limits_{q = 1}^Q {( {\sum\limits_{n = 0}^{{n_p}} {{{(t + {\tau _0} + (q - 1){T_p})}^2} \cdot {{(g(t))}^2}} } )\biggr{|}_{t=n \Delta}{{\left| {{b_q}} \right|}^2}} }}\right.\nonumber\\
&\left.{{- \frac{a^2P}{{L + a^2P}}\frac{1}{{{E_g}}}\sum\limits_{q = 1}^Q {\gamma _q^2{{\left| {{b_q}} \right|}^2}} }}\right)^{ - 1}.
\end{align}}
The proof is provided in Appendix B} \textcolor[rgb]{0.00,0.00,0.00}{ and it should \textcolor[rgb]{0.00,0.00,0.00}{be noted} that $CR{B_{{\tau _0},b}}=JCR{B_{{\tau _0},b}}$ and $CR{B_{{f_0},b}}=JCR{B_{{f_0},b}}$ which are also shown in  Appendix B.}
%
%
%
%
%\textcolor[rgb]{1.00,0.00,0.00}{It is clear that the effect of increasing either $L$ or $P$ is different because of the impact of the unknown reflection coefficient $a$ (when $a\ne 1$). Actually the factor $ \left( \frac{L+ a^2P}{L P}  \right)|_{P=1} $ captures the exact increase compared to the known signal case (\ref{knsigAFIM}). It is amazing since for $P=1$, $CR{B_{{\tau_0},\bm{s},joint}}$ and $CR{B_{{f_0},\bm{s},joint}}$ approach $CR{B_{{\tau_0},joint}}$ and $CR{B_{{f_0},joint}}$ respectively as $L$ approaches infinity but for $L=1$, $CR{B_{{\tau_0},\bm{s},joint}}$ and $CR{B_{{f_0},\bm{s},joint}}$ approach $a^2CR{B_{{\tau_0},joint}}$ and $a^2CR{B_{{f_0},joint}}$ respectively as $P$ approaches infinity.}
\textcolor[rgb]{0.00,0.00,0.00}{\subsection{Correlated Clutter-Plus-Noise}}
The observations from the direct path in (\ref{refobs}) at the $l$-th look can be \textcolor[rgb]{0.00,0.00,0.00}{collected in a vector} as
\begin{eqnarray}
{{\bm{x}}_{dl}} &=& {[{x_{dl}}(0),{x_{dl}}(\Delta),...,{x_{dl}}((N - 1)\Delta)]^T}\nonumber\\
 &=& {\bm s}_0 + \textcolor[rgb]{0.00,0.00,0.00}{{{\bm{w}}_{dl}}}
\end{eqnarray}
where the $N\times 1$ transmitted signal vector ${\bm{{s_0}}}$ is
\begin{align}
{\bm{{s_0}}} = {[s(0),s(\Delta ),...,s((M - 1)\Delta ),0,...,0]^T},
\end{align}
and the $N\times 1$ noise vector \textcolor[rgb]{0.00,0.00,0.00}{${{\bm{w}}_{dl}}$} at the $l$-th look is
\begin{align}\label{noise}
 \textcolor[rgb]{0.00,0.00,0.00}{{{\bm{w}}_{dl}}}=[w_{dl}(0),w_{dl}(\Delta),...,w_{dl}((N-1)\Delta)]^T.
 \end{align}
Similarly, the observations from the reflected path in (\ref{changeobs}) at the $l$-th look can be collected in a vector as
\begin{eqnarray}\label{recvet}
\begin{array}{l}
{{\bm{x}}_{rl}} = {[{x_{rl}}(0),{x_{rl}}(\Delta),...,{x_{rl}}((N - 1)\Delta)]^T}
 = {\bm s}_{\tau_0 f_0} + \textcolor[rgb]{0.00,0.00,0.00}{{{\bm{w}}_{rl}}}
\end{array}
\end{eqnarray}
where the $N\times 1$ transmitted signal vector ${\bm s}_{\tau_0 f_0}$ is
\begin{eqnarray}
{{\bm{s}}_{{\tau _0}{f_0}}} = [0,...,0,s(0){e^{j2\pi {f_0}{n_0}\Delta }},s(\Delta){e^{j2\pi {f_0}{(n_0+1)}\Delta }},...,\nonumber\\s((M - 1)\Delta){e^{j2\pi {f_0}({n_0} + M - 1)\Delta }},0,...,0]^T,
\end{eqnarray}
and the $N\times 1$ noise vector \textcolor[rgb]{0.00,0.00,0.00}{${{\bm{w}}_{rl}}$} at the $l$-th look is
\begin{align}
 \textcolor[rgb]{0.00,0.00,0.00}{{{\bm{w}}_{rl}}}=[w_{rl}(0),w_{rl}(\Delta),...,w_{rl}((N-1)\Delta)]^T.
 \end{align}

The observations from the direct path and reflected path can be written as
\begin{eqnarray}\label{vectorsignal}
{\bm x} = {[{\bm x}_{d1}^T,{\bm x}_{d2}^T,...,{\bm x}_{dL}^T,{\bm x}_{r1}^T,{\bm x}_{r2}^T,...,{\bm x}_{rP}^T]^T}
={\bm s}+{\bm w}
\end{eqnarray}
where the signal vector ${\bm s}$ is
\begin{eqnarray}
{\bm s}={[{\bm s}_{0}^T,{\bm s}_{0}^T,...,{\bm s}_{0}^T,{\bm s}_{\tau_0 f_0}^T,{\bm s}_{\tau_0 f_0}^T,...,{\bm s}_{\tau_0 f_0}^T]^T},
\end{eqnarray}
and the clutter-plus-noise vector ${\bm w}$ is
\begin{eqnarray}
{\bm w}={[{\bm w}_{d1}^T,{\bm w}_{d2}^T,...,{\bm w}_{dL}^T,{\bm w}_{r1}^T,{\bm w}_{r2}^T,...,{\bm w}_{rP}^T]^T}
\end{eqnarray}
which is assumed to be complex Gaussian distributed with zero mean and covariance matrix
${\bm{Q}} = \mathbb {E}\{ {{\bm w}}{{\bm w}^H}\} $. \textcolor[rgb]{0.00,0.00,0.00}{Note that this models either the $a=1$ case or the $a\ne 1$ case since the $a\ne 1$ case in (\ref{Unknowncofreal1}) can be represented as the case in (\ref{recvet}) with a reduction in the noise variance by $\frac{1}{a^2}$. So the $a\ne 1$ case can be handled by modifying ${\bm Q}$. Since we give the results for arbitrary ${\bm Q}$, we already consider the $a\ne 1$ case.}

Using the received signal model previously described but now assuming correlated Gaussian clutter-plus-noise, the pdf of the observation vector is
\begin{equation}
p\left( {{\bm{x}}|{\bm{\theta }}} \right) =
\frac{1}{{{\pi ^{N(P+L)}}\det ( {\bm{C}} )}}\exp ( { -
{{\bm{x}}^H}{{\bm{C}}^{ - 1}}{\bm{x}}} )\label{Prtheta}
\end{equation}
with the covariance matrix  $\bm{C}$ is
\begin{align}
{\bm{C}} &= \mathbb {E}\left\{ {\left( {{\bm{s}} + {\bm{w}}}
\right){{\left( {{\bm{s}} + {\bm{w}}} \right)}^H}} \right\} \notag\\
&= \mathbb {E}\left\{ {{\bm{s}}{{\bm{s}}^H} +
{\bm{w}}{{\bm{w}}^H}} \right\} \notag\\ &=  \mathbb {E}\{ {\bm{s}}{{\bm{s}}^H}\}+
{\bm{Q}}\label{C}.
\end{align}
%Then we reformulate the covariance matrix $\bm{C}$ in a somewhat more explicit matrix form
%\begin{align}\label{CFC}
%{\bm{C}} = {\left[ {\begin{array}{*{20}{c}}
%{{{\bm{1}}_{L \times L}} \otimes {{\bm{B}}_{2M \times 2M}}}&{{{\bm{1}}_{L \times P}} \otimes {{\bm{D}}_{2M \times 2M}}}\\
%{{{({{\bm{1}}_{L \times P}} \otimes {{\bm{D}}_{2M \times 2M}})}^H}}&{{{\bm{1}}_{P \times P}} \otimes {{\bm{F}}_{2M \times 2M}}}
%\end{array}} \right]}
%\end{align}
%with
%\begin{align}
%{{\bm{B}}_{2M \times 2M}} = {{\bm{s}}_0} \cdot {\bm{s}}_0^H,
%\end{align}
%\begin{align}
%{{\bm{D}}_{2M \times 2M}} = {{\bm{s}}_0} \cdot {\bm{s}}_{{\tau _0}{f_0}}^H,
%\end{align}
%\begin{align}
%{{\bm{F}}_{2M \times 2M}} = {{\bm{s}}_{{\tau _0}{f_0}}} \cdot {\bm{s}}_{{\tau _0}{f_0}}^H.
%\end{align}

%The log-likelihood function
%can be written as
%\begin{align}
%  L\left( {{\bm{r}}|{\bm{\theta }}} \right)
%  &= \ln p\left( {{\bm{r}}|{\bm{\theta }}} \right)\notag \\
%   &=  - {{\bm{r}}^H}{{\bm{C}}^{ - 1}}{\bm{r}} - \ln \left( {\det
%   \left( {\bm{C}} \right)} \right) - N(P+L)\ln \left( \pi  \right)\label{logFun}.
%\end{align}

The $(i,j)$th element of
the FIM for the parameter vector ${\bm\theta} = (  \tau_0, f_0,
{s_R}(0), {s_I}(0), {s_R}(\Delta), \ldots, {s_I}(\Delta(M-1))^T$ is given by \cite{kay}
\begin{align}
  {\left[ {\bm{I}(\bm{\theta})} \right]_{i,j}} =
  \text{Tr}\left( {{{\bm{C}}^{ - 1}}\frac{{\partial {\bm{C}}}}{{\partial {{\theta} _i}}}{{\bm{C}}^{ - 1}}\frac{{\partial {\bm{C}}}}{{\partial {{\theta} _j}}}} \right)\label{J}.
\end{align}
We can rewrite (\ref{J}) as \cite{Generalized:2016}
\begin{align}
  {\left[ {{\bm{I}}\left( \bm \theta  \right)} \right]_{i,j}} =&Tr\left( {\frac{{\partial {\bm{C}}}}{{\partial {\theta _i}}}{{\bm{C}}^{ - 1}}\frac{{\partial {\bm{C}}}}{{\partial {\theta _j}}}{{\bm{C}}^{ - 1}}} \right) \notag\hfill \\
   =& {\left( {\frac{{\partial {{\bm{C}}_{vec}}}}{{\partial {\theta _i}}}} \right)^H}\left( {{{\bm{C}}^{ - \dag}} \otimes {{\bm{C}}^{ - 1}}} \right)\left( {\frac{{\partial {{\bm{C}}_{vec}}}}{{\partial {\theta _j}}}} \right)\label{Jvarij}
\end{align}
where ${{\bm{C}}_{vec}}= {\text{vec}}\left( {\bm{C}}
\right)\label{Cvec}$.
Then the FIM for estimating ${\bm\theta}$ is
\begin{align}\label{Icorrelated}
{\bm{I}}(\bm\theta ) = \left[ {\begin{array}{*{20}{c}}
{{J_{{\tau _0}{\tau _0}}}}&{{J_{{\tau _0}{f_0}}}}&{{{\bm{J}}_{{\tau _0}{{\bm {s}}_a}}}}\\
{{J_{{\tau _0}{f_0}}}}&{{J_{{f_0}{f_0}}}}&{{{\bm{J}}_{{f_0}{{\bm {s}}_a}}}}\\
{{{\bm{J}}_{{{\bm {s}}_a}{\tau _0}}}}&{{{\bm{J}}_{{{\bm {s}}_a}{f_0}}}}&{{{\bm{J}}_{{{\bm {s}}_a}{{\bm {s}}_a}}}}
\end{array}} \right]
\end{align}
where ${{ {J}}_{{{\tau_0 \tau_0 }}}} = {\bm{J}}_{{\tau_0
}}^H{{\bm{J}}_{{\tau_0 }}}$, ${{ {J}}_{{{\tau_0 f_0}}}}
={{\bm{J}}_{{ {f_0 \tau_0}}}^H}= {\bm{J}}_{ {\tau_0
}}^H{{\bm{J}}_{ {f_0}}}$, ${{\bm{J}}_{{{\tau_0
}}{{{\bm {s}}_a}}}} = {{\bm{J}}_{{{{\bm {s}}_a}}{ {\tau_0
}}}^H}={ {J}}_{ {\tau_0 }}^H{{\bm{J}}_{{{{\bm {s}}_a}}}}$, ${{ {J}}_{{ {f_0 f_0}}}} =
{\bm{J}}_{{ {f_0}}}^H{{\bm{J}}_{ {f_0}}}$, ${{\bm{J}}_{{ {f_0}}{{{\bm {s}}_a}}}}
={{\bm{J}}_{{{{\bm {s}}_a} }{ {f_0}}}^H}=
{\bm{J}}_{ {f_0}}^H{{\bm{J}}_{{{{\bm {s}}_a} }}}$, ${{\bm{J}}_{{{{\bm {s}}_a} }{{{\bm {s}}_a} }}}{\kern 1pt}  =
{\bm{J}}_{{{{\bm {s}}_a} }}^H{{\bm{J}}_{{{{\bm {s}}_a} }}}$,
\begin{align}
{{\bm{J}}_{ {\tau_0 }}} = \left( {{{\bm{C}}^{- \dag/2}} \otimes
{{\bm{C}}^{ - 1/2}}} \right)\frac{{\partial
{{\bm{C}}_{vec}}}}{{\partial {{ {\tau_0 }}}}},
\end{align}
\begin{align}
{{\bm{J}}_{ {f_0}}} = \left( {{{\bm{C}}^{ - \dag/2}} \otimes
{{\bm{C}}^{ - 1/2}}} \right)\frac{{\partial
{{\bm{C}}_{vec}}}}{{\partial {{ {f_0}}}}},
\end{align}
and
\begin{align}\label{Icos}
{{\bm{J}}_{{{\bm{sa}} }}} = \left( {{{\bm{C}}^{ - \dag/2}}
\otimes {{\bm{C}}^{ - 1/2}}} \right)\frac{{\partial
{{\bm{C}}_{vec}}}}{{\partial {{{\bm {s}}_a} }^T}}
\end{align}
with
\begin{align}\label{sa}
\textcolor[rgb]{0.00,0.00,0.00}{{{\bm {s}}_a}} = {[{s_R}(0),{s_I}(0),{s_R}(\Delta ),...,{s_I}((M - 1)\Delta )]^T}.
\end{align}
%with
%\begin{eqnarray}
%{\bm s}={[{\bm s}_{0}^T,{\bm s}_{0}^T,...,{\bm s}_{0}^T,{\bm s}_{\tau_0 f_0}^T,{\bm s}_{\tau_0 f_0}^T,...,{\bm s}_{\tau_0 f_0}^T]^T}
%\end{eqnarray}

Given any unbiased estimator $ \hat{ {\boldsymbol \theta } } $ of an unknown parameter vector
$ {\boldsymbol \theta} $ based on an observation vector $ {\bm x} $, we have \cite{kay}
\begin{equation}
\textup{MSE}=\mathbb{E}\left\{ ( \hat{ {\boldsymbol \theta} } -
{\bm{\theta }} ) ( \hat{ {\boldsymbol \theta} } - {\bm{\theta
}})^T
 \right\} \succeq
 \textup{JCRB}({{\bm{\theta }}}) =
{{\bm{I}}^{ - 1}}({{\bm{\theta }}}).
\label{crb-bound}
\end{equation}
One could calculate the closed form JCRB with unknown signals in correlated Gaussian clutter-plus-noise by using the results from (\ref{C}), (\ref{Icorrelated})--(\ref{sa}).

\subsection{Nonseparated Direct Path and Reflected Path}
In order to simplify our analysis, we only consider real signals and time delay estimation. The received signals with nonseparated direct and reflected path can be represented as
\begin{eqnarray}\label{combrsig}
{x_{l}}(n\Delta ) = s(n\Delta )+s(n\Delta -\tau_0) + {w_l}(n\Delta )
\end{eqnarray}
\textcolor[rgb]{0.00,0.00,0.00}{for $n=0,1,...,N-1$ and $l=1,...,P$} where $s(n\Delta )$ is nonzero only during the duration $0<n<M-1$, and $s(n\Delta-\tau_0 )$ is nonzero only during the duration $n_0<n<n_0+M-1$. The ${w_l}(n\Delta ), n=0,1,...,N-1$ are real clutter-plus-noise samples with variance $\sigma _w^2$ and $P$ is the number of looks at the signal.
The unknown parameter vector is ${\bm\theta} = (  \tau_0, {s}(0), {s}(\Delta), \ldots, {s}((M-1)\Delta))^T$ and the likelihood function of ${\bm{x}}=(x_{10},\ldots,x_{1(N-1)},x_{20},\ldots,x_{P(N-1)})^T$ is
\begin{small}
\begin{align}
p(\bm{x};\bm{\theta}) \propto \exp \{  - \frac{1}{{2\sigma _w^2}}\sum\limits_{l = 1}^{P } {\sum\limits_{n = 0}^{N - 1} {{{( {{x_{l}}[n] -s(n\Delta )- s(n\Delta-\tau_0 )} )}^2} } } \} % MathType!End!2!1!
\end{align}
\end{small}

\begin{figure}[!t]
\centering
\includegraphics{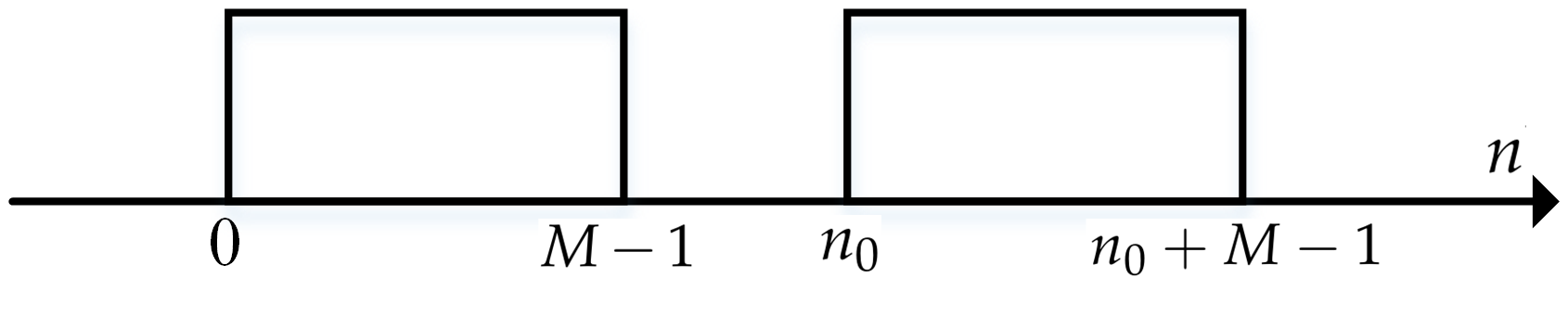}
\caption{A diagram showing the signals from the direct path and the reflected path returns without overlap in time domain.}
\label{nonoverlapshi}
\end{figure}

In the following, the transmitted signal length $M$ is fixed and the impact of overlap on the CRB is investigated by changing the value of $n_0$.  When the two signals from the direct path and reflected path returns do not overlap in time, it implies $n_0 > M-1$, see Fig.~\ref{nonoverlapshi}.
Let $e$ be a scalar, $\bm{b}$ be a $M\times 1$ column vector and $\bm{D}$ be a $M\times M$ matrix. Now the FIM for estimating ${\bm\theta}$ is
\begin{align}\label{FIMreal}
{\bm{I}}(\bm\theta ) = \left[ {\begin{array}{*{20}{c}}
{{{e}}}&{{\bm{b}^T}}\\
{\bm{b}}&{{\bm{D}}}
\end{array}} \right]
\end{align}
with
\begin{align}\label{a}
e= -\mathbb E[\frac{{{\partial ^2}\ln p(\bm{x};\bm\theta)}}{{\partial \tau _0^2}}] =\frac{{P}}{{\sigma _w^2}}\sum\limits_{n = 0}^{M - 1} {{{( {\frac{{\partial s(t)}}{{\partial {t}}}} )^2}}}\biggr{|}_{t=n \Delta},
\end{align}
\begin{align}\label{I1j}
{\bm{b}}_{j} = -\mathbb E[\frac{{{\partial ^2}\ln p(\bm{x};\bm\theta)}}{{\partial \tau _0}{{\partial s(n\Delta)}}}] = - \frac{P}{{\sigma _w^2}}\frac{{\partial s(t)}}{{\partial t}}\biggr{|}_{t=n \Delta}\ &\mbox{if $ j=n+1$},
\end{align}
\begin{align}\label{Ijj}
{\bm{D}{_{j,j}}} = -\mathbb E[\frac{{{\partial ^2}\ln p(\bm{x};\bm\theta)}}{{\partial s(n\Delta)}{{\partial s(n\Delta)}}}] =\frac{{2P}}{{\sigma _w^2}}\quad &\mbox{if $ j=n+1$},
\end{align}
\textcolor[rgb]{0.00,0.00,0.00}{for $n = 0,\ldots, M-1$} and the other FIM entries, not mentioned, are all zero. The CRB with nonoverlapped signals satisfies
 \begin{align}\label{tau_0,non}
 CRB_{\tau_0,non} &= \left[ {\bm{I}}(\bm\theta )\right]^{-1}_{11} =  \left(  e - \sum_{j=1}^{M} {\bm{b}}_{j} {\bm{D}{_{j,j}^{-1}}} {\bm{b}}_{j} \right)^{-1}\nonumber\\
 &= \frac{2P}{P^2}\frac{{\sigma _w^2}}{{\sum\limits_{n = 0}^{M - 1} {{{\left( {\frac{{ds(t)}}{{dt}}} \right)}^2}\biggr{|}_{t=n \Delta}} }}
 \end{align}
 which follows, as expected, the previous results we gave before when we set $L=P$ but now we only estimate the time delay and signal samples without Doppler shift estimation.

\begin{figure}[!t]
\centering
\includegraphics{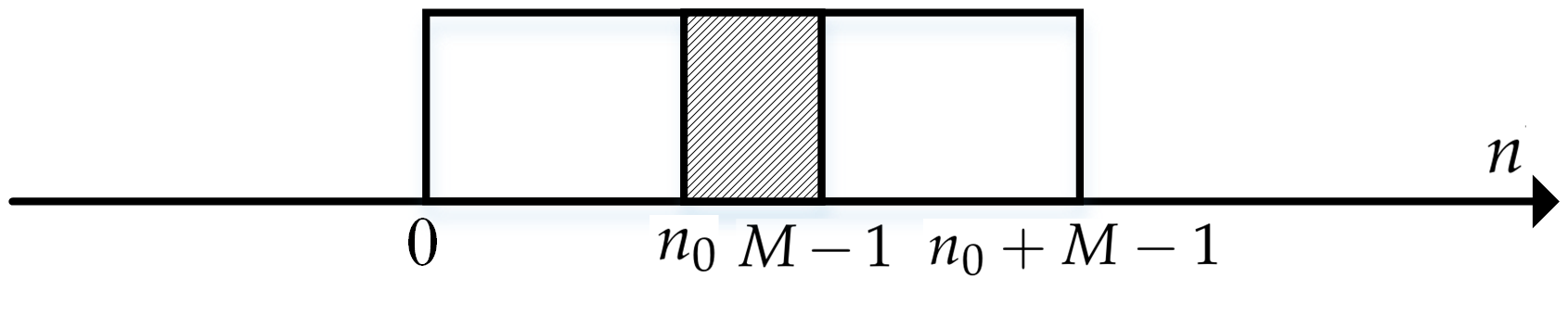}
\caption{A diagram showing the signals from the direct path and the reflected path returns with overlap in time domain.}
\label{overlapshi}
\end{figure}

When the two signals from the direct path and reflected path returns are overlapped in time, it implies \textcolor[rgb]{0.00,0.00,0.00}{$0\le n_0 \le M-1$}, see Fig.~\ref{overlapshi}.

When these two signals are totally overlapped, which implies $n_0=0$. Let $e$ be a scalar defined in (\ref{a}), $\bm{b'}$ be a $M\times 1$ column vector and $\bm{D'}$ be a $M\times M$ matrix. The FIM  for estimating ${\bm\theta}$ becomes
\begin{align}\label{FIMrealtotalover}
{\bm{I}}(\bm\theta ) = \left[ {\begin{array}{*{20}{c}}
{{{e}}}&{{\bm{b'}^T}}\\
{\bm{b'}}&{{\bm{D}}'}
\end{array}} \right]
\end{align}
with
\begin{align}\label{I1jtotalover}
{\bm{b'}}_{j} = -\mathbb E[\frac{{{\partial ^2}\ln p(\bm{x};\bm\theta)}}{{\partial \tau _0}{{\partial s(n\Delta)}}}] = - \frac{2P}{{\sigma _w^2}}\frac{{\partial s(t)}}{{\partial t}}\biggr{|}_{t=n \Delta}\ &\mbox{if $ j=n+1$},
\end{align}
\begin{align}\label{Ijjtotalover}
{\bm{D'}{_{j,j}}} = -\mathbb E[\frac{{{\partial ^2}\ln p(\bm{x};\bm\theta)}}{{\partial s(n\Delta)}{{\partial s(n\Delta)}}}] =\frac{{4P}}{{\sigma _w^2}}\quad &\mbox{if $ j=n+1$},
\end{align}
\textcolor[rgb]{0.00,0.00,0.00}{for $n = 0,\ldots, M-1$}, and the other FIM entries, not mentioned, are all zero. Now
\begin{align}
\frac{1}{\left[ {\bm{I}}(\bm\theta )\right]^{-1}_{11}}&=e - \sum_{j=1}^{M} {\bm{b'}}_{j} {\bm{D'}{_{j,j}^{-1}}} {\bm{b'}}_{j}\nonumber\\
&= \frac{{P}}{{\sigma _w^2}}\sum\limits_{n = 0}^{M - 1} {{{( {\frac{{\partial s(t)}}{{\partial {t}}}} )^2}}}\biggr{|}_{t=n \Delta} - \frac{\sigma _w^2}{4P} \sum\limits_{n = 0}^{M - 1} {{{ \frac{4P^2}{\sigma _w^4} ( {\frac{{\partial s(t)}}{{\partial {t}}}} )^2}}}\biggr{|}_{t=n \Delta} \nonumber\\
&= 0
\end{align}
%and by using Guttman rank additivity formula, we know
%\begin{align}
%rank({\bm{I}}(\bm\theta ))= rank(\bm{D'}) + rank(a - \sum_{j=1}^{M} {\bm{b'}}_{1,j} {\bm{D'}{_{j,j}^{-1}}} {\bm{b'}}_{j,1}^T) = M < M+1
%\end{align}
which means the CRB does not exist in this case.

\textcolor[rgb]{0.00,0.00,0.00}{Assume the} two signals are partially overlapped, which implies $0<n_0\le M-1$. Let $e$ be a scalar defined in (\ref{a}), $\bm{b''}$ be a $M\times 1$ column vector and $\bm{D''}$ be a $M\times M$ matrix. The FIM  for estimating ${\bm\theta}$ is
\begin{align}\label{FIMrealtotalover}
{\bm{I}}(\bm\theta ) = \left[ {\begin{array}{*{20}{c}}
{{{e}}}&{{\bm{b''}^T}}\\
{\bm{b''}}&{{\bm{D}}''}
\end{array}} \right]
\end{align}
with
%where the specific elements can be divided into two parts, the same value of elements in (\ref{I11})--(\ref{Ijj}) plus some different elements due to the overlap. Specifically, $a$, $\bm{D''}{_{j,j}}$ have the same expressions as the corresponding elements in (\ref{I11}) and (\ref{Ijj}). Besides, when the signal samples belong to the non-overlapped part of the signal which implies $n=0,...,n_0-1$, the entries like $\bm{b''}{_{1,j}}$ is the same as those in (\ref{I1j}) where signal samples are all from reflected path. However, when the signal samples belong to the overlapped part, it implies when $n=n_0,...,M-1$, the terms will be different. First, we obtain the elements providing the information about the time delay corresponding to the signal samples from direct path and reflected path,
\begin{small}
\begin{align}
&{\bm{b''}}_{j} = -\mathbb E[\frac{{{\partial ^2}\ln p(\bm{x};\bm\theta)}}{{\partial \tau _0}{{\partial s(n\Delta)}}}] \nonumber\\
&= \left\{ {\begin{array}{*{20}{c}}
- \frac{P}{{\sigma _w^2}}\frac{{\partial s(t)}}{{\partial t}}\biggr{|}_{t=n \Delta}\ &\mbox{if $ j=n+1$}, n = 0,\ldots, n_0-1\\
- \frac{P}{{\sigma _w^2}}\left[\frac{{\partial s(t_1)}}{{\partial t_1}}+ \frac{{\partial s(t_2)}}{{\partial t_2}}\right]\biggr{|}_{t_1=t_2+\tau_0=n \Delta}\ &\mbox{if $ j=n+1$}, n = n_0,\ldots, M-1,
\end{array}} \right.
\end{align}
\end{small}
and
\begin{small}
\begin{align}
&\bm{D''}{_{j,j'}} =\bm{D''}{_{j',j}}\nonumber\\
&= \left\{ {\begin{array}{*{20}{c}}
-\mathbb E[\frac{{{\partial ^2}\ln p(\bm{x};\bm\theta)}}{{\partial s(n\Delta)}{{\partial s(n\Delta)}}}] =\frac{{2P}}{{\sigma _w^2}} &\mbox{if $ j=j'=n+1$}, n = 0,\ldots, M-1\\
-\mathbb E\left[ {\frac{{{\partial ^2}\ln p}}{{\partial {s}[n]\partial {s}[n - {n _0}]}}} \right]
 = \frac{{P}}{{\sigma _w^2}} &\mbox{if $j=n+1, j'=j-n_0$}, n = n_0,\ldots, M-1\\
0\qquad &\mbox{elsewhere}.
\end{array}} \right.
\end{align}
\end{small}

Next since it is complicated to get the inverse of $\bm{D''}$ in (\ref{FIMrealtotalover}) when ${n_0} \in [1,\frac{M}{2})$, we consider the CRB with the partially overlapped signals in the special range  where ${n_0} \in [\frac{M}{2},M - 1]$.

When $n_0 =\frac{M}{2}$, the inverse of $\bm{D''}$ in (\ref{FIMrealtotalover}) becomes
\begin{align}
\bm{D''}{^{ - 1}} =\frac{{\sigma _w^2}}{{P}} \left[ {\begin{array}{*{20}{c}}
{\frac{2}{3}{{\bm{1}}_{{n_0} \times {n_0}}}}&{ - \frac{1}{3}{{\bm{1}}_{{n_0} \times {n_0}}}}\\
{ - \frac{1}{3}{{\bm{1}}_{{n_0} \times {n_0}}}}&{\frac{2}{3}{{\bm{1}}_{{n_0} \times {n_0}}}}
\end{array}} \right],
\end{align}
and the CRB with overlapped signals satisfies
 \begin{align}\label{tau_0,overlap1}
 CRB_{\tau_0,overlap} &= \left(  e - {{\bm{b''}^T}}\bm{D''}{^{ - 1}}{\bm{b''}} \right)^{-1}\nonumber\\
 &= \frac{{\sigma _w^2}}{P}{\left[\frac{1}{3} {\sum\limits_{n = n_0}^{M - 1} {{{\left( {\frac{{ds({t_1})}}{{d{t_1}}} - \frac{{ds({t_2})}}{{d{t_2}}}} \right)}^2}} } \right]^{ - 1}}\biggr{|}_{t_1=t_2+\tau_0=n \Delta}.
 \end{align}
%\sout{Note that when ${\frac{{ds({t_1})}}{{d{t_1}}} = \frac{{ds({t_2})}}{{d{t_2}}}}\biggr{|}_{t_1=t_2+\tau_0=n \Delta}$ for all $n=n_0,...,M - 1$, the $CRB_{\tau_0,overlap}$ in (\ref{tau_0,overlap1}) will be undefined.}
%But most of time, ${\frac{{ds({t_1})}}{{d{t_1}}} \ne \frac{{ds({t_2})}}{{d{t_2}}}}\biggr{|}_{t_1=t_2+\tau_0=n \Delta}$ which implies the $CRB_{\tau_0,overlap}$ often does not blow up.

When ${n_0} \in (\frac{M}{2},M - 1]$, the inverse of $\bm{D''}$ in (\ref{FIMrealtotalover}) becomes
\begin{small}
\begin{align}
&\bm{D''}{^{ - 1}} = \nonumber\\
&\frac{{\sigma _w^2}}{P}\left[ {\begin{array}{*{20}{c}}
{\frac{2}{3}{{\bm{1}}_{(M - {n_0}) \times (M - {n_0})}}}&{{{\bm{0}}_{(M - {n_0}) \times (2{n_0} - M)}}}&{ - \frac{1}{3}{{\bm{1}}_{(M - {n_0}) \times (M - {n_0})}}}\\
{{{\bm{0}}_{(2{n_0} - M) \times (M - {n_0})}}}&{\frac{1}{2}{{\bm{1}}_{(2{n_0} - M) \times (2{n_0} - M)}}}&{{{\bm{0}}_{(2{n_0} - M) \times (M - {n_0})}}}\\
{ - \frac{1}{3}{{\bm{1}}_{(M - {n_0}) \times (M - {n_0})}}}&{{{\bm{0}}_{(M - {n_0}) \times (2{n_0} - M)}}}&{\frac{2}{3}{{\bm{1}}_{(M - {n_0}) \times (M - {n_0})}}}
\end{array}} \right]
\end{align}
\end{small}
and the CRB with overlapped signals satisfies
 \begin{align}\label{tau_0,overlap2}
 &CRB_{\tau_0,overlap} =\left(  e - {{\bm{b''}^T}}\bm{D''}{^{ - 1}}{\bm{b''}} \right)^{-1}\nonumber\\
 &= \frac{{\sigma _w^2}}{P}{\left[ {\frac{1}{3}\sum\limits_{n = {n_0}}^{M - 1} {{{\left( {\frac{{ds({t_1})}}{{d{t_1}}} - \frac{{ds({t_2})}}{{d{t_2}}}} \right)}^2}\biggr{|}_{t_1=t_2+\tau_0=n \Delta}  }}\right.}\nonumber\\
 &\qquad\qquad{\left.{{+  \frac{1}{2}\sum\limits_{n = M - {n_0}}^{{n_0} - 1} {{{\left( {\frac{{ds({t_3})}}{{d{t_3}}}} \right)}^2}\biggr{|}_{t_3=n \Delta}} } } \right]^{ - 1}}.
 \end{align}

\begin{figure}[!t]
\centering
\includegraphics[width=3.0in]{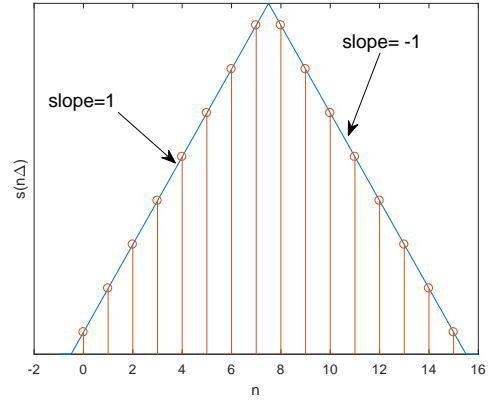}
\caption{A diagram showing the triangle wave with \textcolor[rgb]{0.00,0.00,0.00}{$M=16$}.}
\label{triangle}
\end{figure}

Next, we give an example of a triangle wave, see Fig.~\ref{triangle} which shows
\begin{align}
\frac{{ds(t)}}{{dt}}\biggr{|}_{t=0} =...= \frac{{ds(t)}}{{dt}}\biggr{|}_{t=7\Delta} = -\frac{{ds(t)}}{{dt}}\biggr{|}_{t=8\Delta}=... = -\frac{{ds(t)}}{{dt}}\biggr{|}_{t=15\Delta} = 1.
 \end{align}
By using the above results in (\ref{tau_0,non}) (\ref{tau_0,overlap1}) (\ref{tau_0,overlap2}), we have
 \begin{align}\label{exampleCRB}
CRB_{\tau_0,non}= \frac{{\sigma _w^2}}{{P}}\frac{2}{{M }}\quad &\mbox{when $n_0>M-1$},
  \end{align}
and
\begin{align}\label{zhijiao}
CRB_{\tau_0,overlap} = \frac{{\sigma _w^2}}{{P}}\frac{6}{{5M - 2{n_0}}}\ &\mbox{when $\frac{M}{2}\le n_0 \le M-1$}.
 \end{align}
The derivative of $CRB_{\tau_0,overlap}$ in (\ref{zhijiao}) is
\begin{align}\label{zhijiaolap}
\frac{{\partial CRB_{\tau_0,overlap}}}{{\partial {n_0}}} =  \frac{{12\sigma _w^2}}{{P{{(5M-2{n_0} )}^2}}}
  \end{align}
which is always positive. \textcolor[rgb]{0.00,0.00,0.00}{When $\frac{M}{2}\le n_0 \le M-1$ for $M>1$, then as $n_0$ increases, the size of overlap decreases and the $CRB_{\tau_0,overlap}$ increases towards its maximum of $\frac{{\sigma _w^2}}{P}\frac{6}{(3M+2)}$ but is always smaller than $CRB_{\tau_0,non}$ in (\ref{exampleCRB}). Note that the CRBs in (\ref{zhijiao}) and (\ref{zhijiaolap}) are finite.} The analysis for general signals seems difficult.
\subsection{Multipath}
Again consider the case where the direct path and reflected path returns are separated and the clutter-plus-noise is uncorrelated. It should be noting that in general, the effects of multipath propagation can be modeled by using a linear time-varying channel filter \cite{Viswanath}. The observations from the direct path can be modeled as ($*$ denotes convolution)
\begin{equation}\label{channeld}
{x_{dl}}(n\Delta ) = h(n\Delta ) * s(n\Delta ) + \textcolor[rgb]{0.00,0.00,0.00}{{w_{dl}}(n\Delta )}
\end{equation}
\textcolor[rgb]{0.00,0.00,0.00}{for $n = 0,1,\ldots, N-1$ and $\ell = 1, \ldots, L$}. The observations from the reflected path can be represented as
\begin{equation}\label{channelr}
{x_{rl}}(n\Delta ) = h(n\Delta ) * s(n\Delta  - {\tau _0}){e^{j2\pi {f_0}n\Delta }} + \textcolor[rgb]{0.00,0.00,0.00}{{w_{rl}}(n\Delta )}
\end{equation}
\textcolor[rgb]{0.00,0.00,0.00}{for $n = 0,1,\ldots, N-1$ and $\ell = 1, \ldots, P$}, where $h$ is denoted as the channel filter. If the multipath channel $h$ is known, we can jointly estimate the time delay and Doppler shift using the same method as in Section \uppercase\expandafter{\romannumeral 3}. If the multipath channel $h$ is unknown, we can jointly estimate the time delay and Doppler shift \textcolor[rgb]{0.00,0.00,0.00}{and $h$} using a similar approach as shown in Section \uppercase\expandafter{\romannumeral 3}.

\section{Conclusions}

The CRB for joint time delay and Doppler shift estimation with unknown signals with either unknown or known structure was derived in this paper. The relationship between the CRB of unknown signals and that of known signals has been developed. The theoretical analysis and numerical results show that increasing the number of looks from the direct path and the reflected path returns can help us achieve \textcolor[rgb]{0.00,0.00,0.00}{the} specified estimation performance. \textcolor[rgb]{0.00,0.00,0.00}{The} advantages of known signal format with unknown parameters over totally unknown signals \textcolor[rgb]{0.00,0.00,0.00}{have been shown}. \textcolor[rgb]{0.00,0.00,0.00}{After analysis under a simple common  SCNR model with separated direct and reflected path signals, white clutter-plus-noise and line of sight propagation, extensions to cases with different direct and reflected path SCNRs, correlated clutter-plus-noise, nonseparated direct and reflected path signals and multipath propagation have been discussed.} These results generalize previous
results for a known transmitted signal and describe the number of looks needed to obtain accurate estimation in the asymptotic region where the CRB tightly bounds good estimators.  Extensions to other lower bounds with less restrictions would be a logical next step, but the CRB, being the simplest and most studied lower bound, seems a proper first step in this new direction.

\appendices
%-------------------------- See Appendix \ref{app:computB}.
\textcolor[rgb]{0.00,0.00,0.00}{{\section{Proof of (\ref{tauknown}) and (\ref{fknown}) }}}
\textcolor[rgb]{0.00,0.00,0.00}{Note that the $(i,j)$th entry of the FIM in this multiple parameter case can be computed as (\ref{FIMkay}). Using our previously defined notation,}
\begin{eqnarray}
\ln p(\bm{x};\bm{\theta}) \propto \frac{-1}{\sigma_w^2} \sum_{n=0}^{N-1} \Big| x(n\Delta) - s(n\Delta - \tau_0)e^{j2\pi f_0 n\Delta}\Big|^2.
\end{eqnarray}
The specific entries in the $2\times 2$ FIM for estimating ${\bm\theta}=(\tau_0,f_0)^T$ in this case are
\begin{small}
\begin{align}\label{knowI11}
{\bm I}{(\bm\theta )_{1,1}} = -\mathbb E[\frac{{{\partial ^2}\ln p(\bm{x};\bm\theta)}}{{\partial \tau _0^2}}] =\frac{{2}}{{\sigma _w^2}}\sum\limits_{n = 0}^{M - 1} {{{\left| {\frac{{\partial s(t)}}{{\partial {t}}}} \right|}^2}}\biggr{|}_{t=n \Delta},
\end{align}
\end{small}
\begin{small}
\begin{align}
{\bm I}{(\bm\theta )_{2,2}}=  -\mathbb E[\frac{{{\partial ^2}\ln p(\bm{x};\bm\theta)}}{{\partial f_0^2}}] =\frac{{8{\pi ^2}}}{{\sigma _w^2}}\sum\limits_{n = 0}^{M - 1} {{{(t+\tau_0)}^2}{{\left| {s(t)} \right|}^2}}\biggr{|}_{t=n \Delta},
\end{align}
\end{small}
and
\begin{small}
\begin{align}
{\bm I}{(\bm\theta )_{2,1}}={\bm I}{(\bm\theta )_{1,2}}&=-\mathbb E[\frac{{{\partial ^2}\ln p(\bm{x};\bm\theta)}}{{\partial {\tau _0}\partial {f_0}}}]=\frac{{4\pi}}{{\sigma _w^2}}\sum\limits_{n = 0}^{M - 1} {(t + {\tau _0})}\nonumber\\
&({s_I}(t)\frac{{\partial {s_R}(t)}}{{\partial t}} - {s_R}(t)\frac{{\partial {s_I}(t)}}{{\partial t}})\biggr{|}_{t=n \Delta}.
\end{align}
\end{small}
Further,
\begin{small}
\begin{align}
JCR{B_{{\tau_0}}}&= \left[I(\bm\theta )^{ - 1}\right]_{1,1} = \frac{{\bm I}{(\bm\theta )_{2,2}}}{  {\bm I}{(\bm\theta )_{1,1}}{\bm I}{(\bm\theta )_{2,2}}-{\bm I}{(\bm\theta )_{1,2}}{\bm I}{(\bm\theta )_{2,1}}},
\end{align}
\end{small}
and
\begin{small}
\begin{align}\label{knoJCRBf}
JCR{B_{{f_0}}}&= \left[I(\bm\theta )^{ - 1}\right]_{2,2} = \frac{{\bm I}{(\bm\theta )_{1,1}}}{{\bm I}{(\bm\theta )_{1,1}}{\bm I}{(\bm\theta )_{2,2}}-{\bm I}{(\bm\theta )_{1,2}}{\bm I}{(\bm\theta )_{2,1}}}.
\end{align}
\end{small}Using (\ref{knowI11})--(\ref{knoJCRBf}), the specific expressions of $JCR{B_{{\tau_0}}}$ and $JCR{B_{{f_0}}}$ are shown in (\ref{tauknown}) and (\ref{fknown}).

\textcolor[rgb]{0.00,0.00,0.00}{\section{Proof of (\ref{aJCRBtaub})--(\ref{aCRBfb})} }\label{app:computB}
Using our previously defined notation, the FIM for estimating ${\bm\theta}_{1} = (\tau_0,f_0, a)^T$ with known signals from (\ref{Unknowncofreal1}) is given in (\ref{knsigAFIM}), \textcolor[rgb]{0.00,0.00,0.00}{shown at the top of the next page.}
\begin{figure*}[!t]
%\normalsize
\begin{small}
\begin{align}\label{knsigAFIM}
{\bm{I}_{ks}}({\bm\theta}_{1} ) = \left[ {\begin{array}{*{20}{c}}
{\frac{{2{a^2}}}{{\sigma _w^2}}\sum\limits_{n = 0}^{M - 1} {{{\left| {\frac{{\partial s(t)}}{{\partial t}}} \right|}^2}} }{\biggr{|}_{t=n \Delta}}&{\frac{{4\pi {a^2}}}{{\sigma _w^2}}\sum\limits_{n = 0}^{M - 1} {\left( {t  + {\tau _0}} \right)\left( {{s_I}(t)\frac{{\partial {s_R}(t)}}{{\partial t}} - {s_R}(t)\frac{{\partial {s_I}(t)}}{{\partial t}}} \right)} }{\biggr{|}_{t=n \Delta}}&{\frac{{ - 2{a}}}{{\sigma _w^2}}\sum\limits_{n = 0}^{M - 1} {\left( {{s_R}(t)\frac{{\partial {s_R}(t)}}{{\partial t}} + {s_I}(t)\frac{{\partial {s_I}(t)}}{{\partial t}}} \right)} }{\biggr{|}_{t=n \Delta}}\\
{\frac{{4\pi {a^2}}}{{\sigma _w^2}}\sum\limits_{n = 0}^{M - 1} {\left( {t  + {\tau _0}} \right)\left( {{s_I}(t)\frac{{\partial {s_R}(t)}}{{\partial t}} - {s_R}(t)\frac{{\partial {s_I}(t)}}{{\partial t}}} \right)} }{\biggr{|}_{t=n \Delta}}&{\frac{{8{\pi ^2}{a^2}}}{{\sigma _w^2}}\sum\limits_{n = 0}^{M - 1} {{{\left( {t  + {\tau _0}} \right)}^2}{{\left| s(t) \right|}^2}} }{\biggr{|}_{t=n \Delta}}&0\\
{\frac{{ - 2{a}}}{{\sigma _w^2}}\sum\limits_{n = 0}^{M - 1} {\left( {{s_R}(t)\frac{{\partial {s_R}(t)}}{{\partial t}} + {s_I}(t)\frac{{\partial {s_I}(t)}}{{\partial t}}} \right)} }{\biggr{|}_{t=n \Delta}}&0&{\frac{2}{{\sigma _w^2}}\sum\limits_{n = 0}^{M - 1} {{{\left| s(t) \right|}^2}} }{\biggr{|}_{t=n \Delta}}
\end{array}} \right]
\end{align}
\end{small}
\hrulefill
\vspace*{-1em}
\end{figure*}
It is worth noting that the JCRBs with known signals, namely $JCR{B_{{\tau_0}}}$, $JCR{B_{{f_0}}}$ and $JCR{B_{{a}}}$ are the diagonal entries in ${\bm{I}_{ks}}({\bm\theta}_{1})^{-1}$ \textcolor[rgb]{0.00,0.00,0.00}{and they are calculated with only one look which is discussed at the end of Section \uppercase\expandafter{\romannumeral 3}}.

For unknown signals with multiple looks, the entries in the FIM in (\ref{FIMbrief}) for estimating ${\bm\theta} = (  \tau_0, f_0,a,
{s_R}(0),{s_I}(0), {s_R}(\Delta),\\ \ldots, {s_I}((M-1)\Delta)^T$ are
\begin{align}
\label{APIks}{\bm{A}}=P{\bm{I}_{ks}},\\
\label{g}\bm{C}_{j,j} = {\frac{{2L + 2{a^2}P}}{{\sigma _w^2}}} \qquad &\mbox{if $j=1,2,...,2M$},
\end{align}
and ${\bm{B}}$ is given in (\ref{f}), \textcolor[rgb]{0.00,0.00,0.00}{shown at the top of the next page}.

\begin{figure*}[!t]
%\normalsize
\begin{small}
\begin{align}\label{f}
{\bm{B}} = \left[ {\begin{array}{*{20}{c}}
{\frac{{ - 2{a^2}P}}{{\sigma _w^2}}\frac{{\partial {s_R}(t)}}{{\partial t}}}{\biggr{|}_{t=0}}&{\frac{{ - 2{a^2}P}}{{\sigma _w^2}}\frac{{\partial {s_I}(t)}}{{\partial t}}}{\biggr{|}_{t=0}}& \cdots &{\frac{{ - 2{a^2}P}}{{\sigma _w^2}}\frac{{\partial {s_I}(t)}}{{\partial t}}}{\biggr{|}_{t=(M-1)\Delta}}\\
{ - \frac{{4\pi {a^2}P}}{{\sigma _w^2}}\left( {t  + {\tau _0}} \right){s_I}(t)}{\biggr{|}_{t=0}}&{\frac{{4\pi {a^2}P}}{{\sigma _w^2}}\left( {t  + {\tau _0}} \right){s_R}(t)}{\biggr{|}_{t=0}}& \cdots &{\frac{{4\pi {a^2}P}}{{\sigma _w^2}}\left( {t  + {\tau _0}} \right){s_R}(t)}{\biggr{|}_{t=(M-1)\Delta}}\\
{\frac{{2aP}}{{\sigma _w^2}}{s_R}(t)}{\biggr{|}_{t=0}}&{\frac{{2aP}}{{\sigma _w^2}}{s_I}(t)}{\biggr{|}_{t=0}}& \cdots &{\frac{{2aP}}{{\sigma _w^2}}{s_I}(t)}{\biggr{|}_{t=(M-1)\Delta}}
\end{array}} \right]
\end{align}
\end{small}
\hrulefill
\vspace*{-1em}
\end{figure*}

Using the expressions of elements in ${{\bm{I}}_{ks}}({\bm\theta}_{1})$, ${\bm{C}}$ and ${\bm{B}}$ derived in (\ref{knsigAFIM}), (\ref{g}) and (\ref{f}), (\ref{tfcontain}) becomes
\begin{align}\label{tfast}
{{\bm A} - \bm{B}{\bm{C}^{ - 1}}{\bm{B}^T}} = \frac{LP}{L+a^2P}{\bm{I}_{ks}}({\bm\theta}_{1}).
\end{align}

\textcolor[rgb]{0.00,0.00,0.00}{Computing the} inverse of ${\bm{I}_{ks}}({\bm\theta}_{1})$ and ${{\bm A} - \bm{B}{\bm{C}^{ - 1}}{\bm{B}^T}}$ in  (\ref{knsigAFIM}) and (\ref{tfast}) respectively, the relationships between the JCRBs with known signals and those with unknown signals are the same as those shown in (\ref{reflecoefftau}) and (\ref{reflecoefff}).
%\textcolor[rgb]{1.00,0.00,0.00}{\begin{align}\label{aunknownJCBTAUA}
%JCR{B_{{\tau_0},\bm{s}}}=\frac{L+a^2P}{LP}JCR{B_{{\tau_0}}},
%\end{align}
%and
%\begin{align}
%JCR{B_{{f_0},\bm{s}}}=\frac{L+a^2P}{LP}JCR{B_{{f_0}}}.
%\end{align}}

If we estimate $\tau_0$ and $f_0$ separately for unknown signals with the unknown factor $a$, we will get the same results as those shown in (\ref{reflecoetau2}), (\ref{reflecoefff2}) and (\ref{reflecoefff2CRB}) but a different (\ref{reflecoetau2CRB}) as follows
\begin{small}
\begin{align}
&CR{B_{{\tau _0}}}= \frac{{\sigma _w^2}}{{2{a^2}}}\cdot\nonumber\\
&\frac{{\sum\limits_{n = 0}^{M - 1} {{{\left| {s(t)} \right|}^2}{\biggr{|}_{t=n \Delta}}} }}{{\sum\limits_{n = 0}^{M - 1} {{{\left| {\frac{{\partial s(t)}}{{\partial t}}} \right|}^2}{\biggr{|}_{t=n \Delta}}} \sum\limits_{n = 0}^{M - 1} {{{\left| {s(t)} \right|}^2}{\biggr{|}_{t=n \Delta}} - } {{\left( {\sum\limits_{n = 0}^{M - 1} {\left( {{s_R}(t)\frac{{\partial {s_R}(t)}}{{\partial t}} + {s_I}(t)\frac{{\partial {s_I}(t)}}{{\partial t}}} \right){\biggr{|}_{t=n \Delta}}} } \right)}^2}}}.
\end{align}
\end{small}

For the known signal structure with unknown parameters, the entries in the FIM in (\ref{lalalFIM}) for estimating ${\bm\theta} = (  \tau_0, f_0,a,
{b_{1R}},{b_{1I}}, {b_{2R}}, \ldots, {b_{QI}})^T$ are
\begin{small}
\begin{align}\label{H}
{\bm{B}'' } = \frac{{2Pa}}{{\sigma _w^2}}\left[ {\begin{array}{*{20}{c}}
{ - a\rho {b_{1R}}}&{ - a\rho {b_{1I}}}& \cdots &{ - a\rho {b_{QI}}}\\
{ - 2\pi a{\gamma _1}{b_{1I}}}&{2\pi a{\gamma _1}{b_{1R}}}& \cdots &{2\pi a{\gamma _Q}{b_{QR}}}\\
{{E_g}{b_{1R}}}&{{E_g}{b_{1I}}}& \cdots &{{E_g}{b_{QI}}}
\end{array}} \right],
\end{align}
\end{small}
and
\begin{eqnarray}\label{U}
\bm{C}''_{j,j} = {\frac{({ 2L + 2{a^2}P })E_g}{{\sigma _w^2}}} \qquad &\mbox{if $j=1,2,...,2Q$}
\end{eqnarray}
with ${\bm A}$ as that shown in (\ref{APIks}). Now ${\bm{V}}={{{({{\bm{A}}} - {{\bm{B}'' }}{({{\bm{C}'' }})^{ - 1}}{\bm{B}''^T})}}}$ becomes a $3\times 3$ matrix and its entries ${\bm{V}}_{1,1}$ and ${\bm{V}}_{2,2}$ in (\ref{f11}) and (\ref{f22}) are
\begin{align}
{\bm{V}}_{1,1} =\frac{{2{a^2}P}}{{\sigma _w^2}}\sum\limits_{q = 1}^Q {{{\left| {{b_q}} \right|}^2}} \left( {\sum\limits_{n = 0}^{{n_p}} {{{\left( {\frac{{dg(t)}}{{dt}}} \right)}^2}\biggr{|}_{t=n \Delta} - \frac{{{a^2}P}}{{L + {a^2}P}}\frac{{{\rho ^2}}}{{{E_g}}}} } \right)
\end{align}
and
\begin{align}
&{\bm{V}}_{2,2}= \frac{{8{\pi ^2}{a^2}P}}{{\sigma _w^2}}\left( \sum\limits_{q = 1}^Q {( {\sum\limits_{n = 0}^{{n_p}} {{{(t + {\tau _0} + (q - 1){T_p})}^2} \cdot {{(g(t))}^2}} } )\biggr{|}_{t=n \Delta}{{\left| {{b_q}} \right|}^2}} \right.\nonumber\\
&\qquad\qquad\left.- \frac{{a^2}P}{L+{a^2}P}\frac{1}{E_g}\sum\limits_{q = 1}^Q {{{\gamma_q ^2}{\left| {{b_q}} \right|}^2}} \right).
\end{align}
But ${\bm{V}}_{1,2}$ and ${\bm{V}}_{2,1}$ are the same as that shown in (\ref{ccc}). Moreover,
\begin{align}
{\bm{V}}_{1,3}&= {\bm{V}}_{3,1}=- \frac{{2{a}P}}{{\sigma _w^2}}\sum\limits_{q = 1}^Q {{{\left| {{b_q}} \right|}^2}}\frac{L}{{L + {a^2}P}}\rho,\\
{\bm{V}}_{3,3}&=\frac{{2P}}{{\sigma _w^2}}\sum\limits_{q = 1}^Q {{{\left| {{b_q}} \right|}^2}} \frac{L}{{L + {a^2}P}}{E_g},
\end{align}
and
\begin{align}
{\bm{V}}_{3,2}&={\bm{V}}_{2,3}={\bm A}_{2,3} - \sum\limits_{i = 1}^{2Q} {{{\bm{B}}''_{2,i}}{{[{{\bm{C''}}^{ - 1}}]}_{i,i}}{{\bm{B}}''_{3,i}}}  = 0.
\end{align}
Then $JCR{B_{{\tau _0},b}}$ and $JCR{B_{{f _0},b}}$ are obtained in (\ref{aJCRBtaub}) and (\ref{aJCRBfb}) by \textcolor[rgb]{0.00,0.00,0.00}{inverting} $\bm V$. It is worth noting that ${\bm{V}}_{1,2}$, ${\bm{V}}_{2,1}$, ${\bm{V}}_{2,3}$ and ${\bm{V}}_{3,2}$ are all zero which means the time delay part of joint estimation will have no effect on the Doppler shift part of joint estimation and vice versa. \textcolor[rgb]{0.00,0.00,0.00}{This means $CR{B_{{\tau _0},b}}=JCR{B_{{\tau _0},b}}$ and $CR{B_{{f_0},b}}=JCR{B_{{f_0},b}}$. We} can show $JCR{B_{{\tau _0},b}}<JCR{B_{{\tau _0},s}}$ by using (\ref{rhorho}). Similarly, we can repeat the calculations to show $JCRB_{f_0,b}<JCRB_{f_0,s}$, $CRB_{\tau_0,b}< CRB_{\tau_0,s}$ and $CRB_{f_0,b}<CRB_{f_0,s}$.

\bibliographystyle{IEEEtran}
\bibliography{refs}
\end{document}